\documentclass[aip,jcp,reprint]{revtex4-1}
\usepackage{graphicx, color}
\usepackage{amsmath, amssymb, amsfonts, mathrsfs}

\begin{document}

\title{A variational formulation of electrostatics in a medium with spatially varying dielectric permittivity}
\author{Vikram Jadhao}
\affiliation{Department of Materials Science and Engineering, Northwestern University, Evanston, Illinois 60208}
\author{Francisco J. Solis}
\affiliation{Department of Integrated Natural Sciences, Arizona State University, Glendale, Arizona 85306}
\author{Monica Olvera de la Cruz}
\email{m-olvera@northwestern.edu}
\affiliation{Department of Materials Science and Engineering, Northwestern University, Evanston, Illinois 60208}
% \date{\today}

\begin{abstract}
In biological and synthetic materials, many important processes involve charges that are 
present in a medium with spatially varying dielectric permittivity. 
To accurately understand the role of electrostatic interactions in such systems, it is important 
to take into account the spatial dependence of the permittivity of the medium. 
However, due to the ensuing theoretical and computational challenges, 
this inhomogeneous dielectric response of the medium is often ignored or excessively simplified.
We develop a variational formulation of electrostatics to accurately investigate 
systems that exhibit this inhomogeneous dielectric response. 
Our formulation is based on a true energy functional of the polarization charge density. 
The defining characteristic of a true energy functional is that at its minimum 
it evaluates to the actual value of the energy; 
this is a feature not found in many commonly used electrostatic functionals. 
We explore in detail the charged systems that exhibit sharp discontinuous change in dielectric
permittivity, and we show that for this case our functional reduces to a 
functional of only the surface polarization charge density.
We apply this reduced functional to study model problems for which analytical solutions are well known. 
We demonstrate, in addition, that the functional has many properties that make 
it ideal for use in molecular dynamics simulations.
\end{abstract}

\maketitle

\section{Introduction}
Many biological systems involve mobile or fixed charges, the electrostatic response of which 
is key to our understanding of the physical behavior of such systems.
Nucleic acids and many proteins are charged in physiologically relevant conditions. 
The interactions that arise from the presence of these charges are crucial in 
the determination of the structure and function of these polymers \cite{honig,perutz}.
Biological processes such as signaling in cells involve the creation of 
electrical potential differences and transport of ions across the cellular membrane \cite{clapham}.
On the other hand, in the design and stabilization of many different synthetic structures, 
the electrostatic forces play a major role.
Examples include self-assembled colloidal dispersions \cite{levin1}, 
polynucleotide adsorption \cite{bedzyk}, DNA precipitation in multivalent salts \cite{rouzina,raspaud}, 
overcharged surfaces \cite{charge_inversion,charge_inversion2}, patterned surfaces \cite{paco}, 
spontaneous adsorption of ions at liquid-liquid interfaces\cite{bier,kung,wang}, faceted thin 
shells \cite{vernizzi}, viral assembly \cite{muthu}, and various dynamical processes including DNA
gel electrophoresis \cite{netz} and 
related polyelectrolyte separation process \cite{depablo,holm}. 
Theoretical investigations of these interesting materials and biological systems 
must therefore accurately incorporate electrostatic interactions. 

Under the conditions of high ionic concentration or in the presence of multivalent ions, 
when the finite size of ions and the inter-ionic correlations become significant, 
mean field theories are generally found inadequate to capture important electrostatic effects \cite{ivan}.
For arbitrarily curved geometries or where the dielectric response of the medium is not homogeneous, 
the associated electrostatics problem gets too complicated even for the more 
sophisticated analytical treatments \cite{lue1}, and the use of numerical techniques becomes necessary.
However, an accurate computer simulation involving electrostatic interactions presents its own challenges.
The first challenge stems from the long range of the Coulomb force which 
implies that every charge interacts with every other charge.
Thus, a system of $N$ charges requires an expensive $O(N^{2})$ force (or energy) 
calculation at every simulation step.
Attempts to ameliorate this scaling behavior have resulted in the development of several methods:
e.g, Ewald summation, particle-mesh methods, fast multipole methods \cite{sagui1}, and  
local electrostatics algorithms \cite{sagui2,maggs-rossetto,rottler-maggs}.
The other main challenge arises due to the presence of dielectric heterogeneities in the medium, 
and this constitutes the main focus of the present paper.

Free charges polarize their surrounding dielectric medium and the resulting net polarization and 
electric fields can have complex behavior.
Modeling of systems with electrostatic interactions should, ideally, 
incorporate this dielectric response of the medium.
An explicit inclusion of the medium components (molecules of the solvent, for example) 
as a part of the model for the real system
renders a prohibitively large number of degrees of freedom to simulate, 
such that even with the most efficient methods \cite{sagui1,sagui2} direct simulation becomes 
too computationally expensive.
In many cases, the introduction of a spatially varying dielectric constant 
in the model is sufficient to capture the effects of polarizability and describe the dielectric response.
In the simplest case of a uniform dielectric response, a single dielectric constant 
can describe a coarse-grained medium, 
and simulations can proceed as they would in free space, albeit with a scaled Coulomb's law.

However, most real situations involve regions with different dielectric response, 
as is the case for proteins within an aqueous cellular medium or 
for emulsions where oil and water are partitioned \cite{sacanna}.
In the presence of this varying dielectric response, the simplest form of  
Coulomb's law breaks down and one has to accurately solve the Poisson equation,
at \emph{each} simulation step, to obtain the necessary force (energy) information for the propagation of ionic coordinates.
This adversely affects the stability and efficiency of the resulting numerical procedure.
Because of these computational challenges, the problem of treating dielectric 
heterogeneities in charge simulations continues to be a subject of 
intense research \cite{marchi,allen1,messina1,boda,attard,maggs-rossetto,linse,gan,tyagi,santos}.

A few previous attempts towards the solution of the problem of inhomogeneous 
dielectric response have involved a reformulation of electrostatics 
as a variational problem \cite{marchi,allen1,attard,maggs-rossetto,wang,lipparini}.
Here, the solution to the Poisson equation is obtained not as a solution of a differential equation, 
but as the extremum of a suitably constructed functional \cite{jackson,schwinger}.
An important advantage of adopting a variational approach is that it offers the possibility of 
bypassing the effort 
to explicitly optimize the functional at each step by framing the
problem in such a way that the very process of updating the simulation guarantees the optimization 
of the functional.
In other words, since the optimization of the functional is equivalent to the solution of the 
Poisson equation, we are offered the possibility of solving the Poisson equation  
\emph{on-the-fly} in tandem 
with the generation of the new charge configuration.
However, this possibility arises only when the variational approach is based 
on an energy functional: a functional which \emph{minimizes} to the \emph{true} electrostatic energy.
We note that in the literature, there is an abundance of functionals 
that are \emph{not} energy functionals \cite{allen1,jackson,radke,karplus,schwinger} 
and therefore the numerical schemes associated with these functionals do not employ 
the ideas of dynamical optimization. 

\begin{figure}[h]
\centerline{
\includegraphics[scale=0.18]{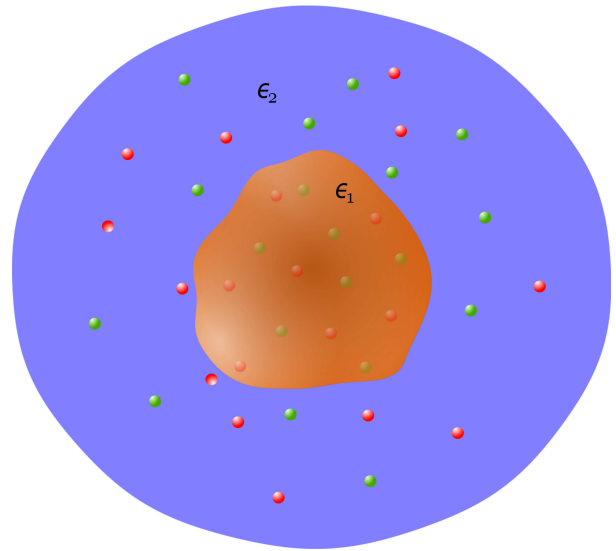}
}
\caption
{\label{fig1}
A uniform dielectric medium characterized by permittivity $\epsilon_{1}$ separated from another region of permittivity $\epsilon_{2}$
by a boundary that is treated as a thin interface.
Positive (red) and negative (green) ions are present in either media. Coarse-grained models of this kind are often employed to study systems such as charged bio-macromolecules in 
aqueous solution or charged colloidal suspensions.
}
\end{figure}

In addition to the need for a true energy functional for an efficient numerical implementation 
of the variational procedure, 
it is also crucial to produce a functional  with the appropriate function variable(s).
In many cases a particular electrostatic variable offers distinct advantage over others.
For example, a coarse-grained model often employed to study phenomena in both biological and synthetic settings 
is that of ions present in piecewise-uniform dielectrics separated by sharp interfaces (see Fig.~\ref{fig1}).
In this case, it is advantageous to choose the polarization charge density as the variable to solve for, 
rather than the electrostatic potential or the polarization vector.
This is because when the dielectric response of the medium is piecewise uniform, 
the unknown polarization charge density resides only on the interface, 
and thus we are presented with the possibility to reduce the full three-dimensional electrostatic 
problem to a two-dimensional one.

In the light of the above discussion, it is clear that
an energy functional of the polarization charge density 
would provide many advantages with regards to the numerical implementation 
of the variational method formulated to study problems involving dielectric heterogeneities. 
In the literature, one finds many functionals along with their associated numerical minimization procedures.
Though a functional of polarization density for a general system has been developed in 
Ref.~\onlinecite{allen1}, it is not an energy functional. 
Similarly, the functionals derived in Refs.~\onlinecite{marcus,felderhof,rottler-maggs} are energy
functionals, but the basic field variables employed are vector fields 
such as the polarization vector \cite{marcus,felderhof} 
or the electric field \cite{rottler-maggs}, not the polarization charge density.
Attard \cite{attard} has provided an energy functional of the surface polarization charge density, 
but this functional is derived for a specific system that involves all 
free charges to be constrained in one uniform dielectric medium.

In a recent paper (Ref.~\onlinecite{shortpaper}), we introduced a variational formulation of 
electrostatics that produced an energy functional of the polarization charge density.
This functional reads:
\begin{equation}\begin{split}\label{eq:fnal}
\mathscr{F}[\omega]&=\frac{1}{2}\iint \rho_{\mathbf{r}}G_{\mathbf{r},\mathbf{r'}}
\left(\rho_{\mathbf{r'}}+\Omega_{\mathbf{r'}}[\omega]\right) d^{3}r'd^{3}r\\
&- \frac{1}{2}\iint \Omega_{\mathbf{r}}[\omega] G_{\mathbf{r},\mathbf{r'}}
\left(\omega_{\mathbf{r'}} - \Omega_{\mathbf{r'}}[\omega] \right) d^{3}r'd^{3}r,
\end{split}\end{equation}
where $\omega$ is the polarization charge density and 
$\Omega_{\mathbf{r}}[\omega]$ is both a functional of $\omega(\mathbf{r})$ and a function of $\mathbf{r}$, and is defined as
\begin{equation}\label{eq:Omega}
\Omega_{\mathbf{r}}[\omega] = \nabla\cdot\left( \chi_{\mathbf{r}}\nabla\int G_{\mathbf{r},\mathbf{r'}}
\left( \rho_{\mathbf{r'}} + \omega_{\mathbf{r'}} \right) d^{3}r'\right).
\end{equation}
In the above equations $\chi$ is the susceptibility of the medium, $\rho$ is the free charge density, 
and $G(\mathbf{r},\mathbf{r'}) = |\mathbf{r} - \mathbf{r'}|^{-1}$ is the Green's function in free space.
In our derivation we assumed that the medium polarization obeys linear response 
and also assumed the existence of Dirichlet boundary conditions.
Both these assumptions are standard from the point of view 
of constructing electrostatic free energy functionals.

$\mathscr{F}[\omega]$ is applicable to any configuration of 
free charges and works for arbitrary spatial variation in dielectric response. 
We also provided the expression of this functional for the important 
case of point charges present in piecewise-uniform dielectrics and 
developed a Car-Parrinello molecular dynamics scheme to study the equilibrium properties 
of such systems. 
As an application, we computed the density profiles for monovalent salt ions 
near a spherical emulsion droplet separating two liquids of different dielectric constants. 

In this paper, we present a detailed derivation of $\mathscr{F}[\omega]$, showing steps 
that were omitted in Ref.~\onlinecite{shortpaper} for the lack of space, and discuss the important features of our 
variational formulation that enable the production of an energy functional.
Also, we explore in detail, with several examples, the particular case of piecewise-uniform dielectric response. 
In addition, we provide the proofs that show that $\mathscr{F}[\omega]$ is an energy functional. 
In the supplementary information of Ref.~\onlinecite{shortpaper} we 
proved the minimum property of this functional at its extremum. 
Here we show that upon extremizing this functional one finds the usual electrostatic 
relation for the polarization charge density and the extremum value of the functional coincides with
the true electrostatic energy of the system.
For the sake of completeness we also include the proof of the functional becoming a minimum at its extremum. 

The paper is organized as follows.
In Sec.~\ref{sec:vform} we derive $\mathscr{F}[\omega]$ and 
discuss the key aspects of our variational formulation.
In Sec.~\ref{sec:pud}, we explicitly specialize $\mathscr{F}[\omega]$ to the 
case of sharp dielectric interfaces, and apply the resulting functional to some simple interfacial 
shapes.
In Sec.~\ref{sec:numeric} we demonstrate a grid-based numerical procedure
to implement the functional minimization and some concluding remarks are made in Sec.~\ref{sec:conclusion}.
In Appendix \ref{sec:ext} we prove that $\mathscr{F}[\omega]$ is an energy functional.
Finally, Appendix \ref{sec:app} discusses the application of our variational principle to the simple 
case of a uniform dielectric.

\section{Variational formulation}\label{sec:vform}
In the first half of this section we provide a detailed derivation of 
the functional given in Eq.~\eqref{eq:fnal}. 
The second half discusses the important features of the variational formulation that determine 
the extremal properties of the resulting functional.
Gaussian units are used throughout.
\subsection{Derivation of $\mathscr{F}[\omega]$}\label{sec:deriv}
We begin with the standard expression for the electrostatic energy 
written in its equivalent functional form:
\begin{equation}\label{eq:fnalinE}
\mathscr{F}\left[\mathbf{E}\right]=\frac{1}{8\pi}\int \epsilon_{\mathbf{r}}\left|\mathbf{E}_{\mathbf{r}}\right|^{2} d^{3}r.
\end{equation}
Here $\epsilon$ is the dielectric permittivity and $\mathbf{E}$ is the electric field.
Following the formulation introduced in Ref. \onlinecite{kung},
we include Gauss's law as a constraint to the functional in \eqref{eq:fnalinE} via the Lagrange multiplier $\phi$, obtaining
\begin{eqnarray}\label{eq:gausslaw}
\mathscr{F}[\mathbf{E},&&\phi]=\frac{1}{8\pi}\int \epsilon_{\mathbf{r}}|\mathbf{E}_{\mathbf{r}}|^{2} d^{3}r \nonumber\\
&&-\int \phi_{\mathbf{r}}\left(\nabla \cdot \left(\frac{\epsilon_{\mathbf{r}}\mathbf{E}_{\mathbf{r}}}{4\pi}\right) -\rho_{\mathbf{r}}\right) d^{3}r.
\end{eqnarray}
We note that $\phi$ can be shown to coincide with the electrostatic potential at equilibrium. 
Also, we take $\mathscr{F}$ to depend parametrically on the free charge density $\rho$, implying that the latter will not be used as a variational field.
We assume that the medium polarization $\mathbf{P}$ obeys linear response: $\mathbf{P}=\chi\mathbf{E}$,
where $\chi$ is the susceptibility connected to $\epsilon$ by the relation $\epsilon = 1 + 4\pi\chi$.
Employing this relation between $\epsilon$ and $\chi$ we now introduce the field variable $\mathbf{P}$ in \eqref{eq:gausslaw} in the following way:
\begin{equation}\label{eq:inP}
\begin{split}
\mathscr{F}[&\mathbf{E},\mathbf{P},\phi]=\frac{1}{8\pi}\int |\mathbf{E}_{\mathbf{r}}|^{2} d^{3}r + \int \frac{|\mathbf{P_{\mathbf{r}}}|^{2}}{2\chi_{\mathbf{r}}} d^{3}r  \\
&- \int \phi_{\mathbf{r}}\left( \nabla \cdot \frac{\mathbf{E}_{\mathbf{r}}}{4\pi} + \nabla \cdot \mathbf{P}_{\mathbf{r}} - \rho_{\mathbf{r}}\right) d^{3}r.
\end{split}
\end{equation}
Variations of \eqref{eq:inP} with respect to $\mathbf{E}$ and $\phi$ give:
\begin{align}
\delta\mathbf{E} : \qquad &\mathbf{E_{\mathbf{r}}} = -\nabla\phi_{\mathbf{r}},\label{eq:ve} \\
\delta\phi : \qquad &\nabla\cdot\mathbf{E_{\mathbf{r}}} = 4\pi\left( \rho_{\mathbf{r}} - \nabla\cdot\mathbf{P}_{\mathbf{r}} \right).\label{eq:vpot}
\end{align}
In obtainig the above variations we make use of the Dirichlet boundary condition (DBC):
\begin{equation}
\phi_{\mathbf{r}} = 0 \quad\textrm{for}\quad \mathbf{r} \in S,
\end{equation}
where $S$ is a boundary invoked at infinity. All the surface integrals that appear as a consequence of taking the variations are rendered void by the use of DBC.
From \eqref{eq:ve} it is clear that $\phi$ must be the electrostatic potential. 
Using \eqref{eq:ve} we eliminate $\mathbf{E}$ from \eqref{eq:vpot} and obtain
\begin{equation}\label{eq:modpoisson}
\nabla^{2}\phi_{\mathbf{r}} = -4\pi\left( \rho_{\mathbf{r}} - \nabla\cdot\mathbf{P}_{\mathbf{r}} \right).
\end{equation}
Eq.~\eqref{eq:modpoisson} is the Poisson equation satisfied by the potential $\phi$ when the charge density in \emph{free space} is given by $\rho - \nabla\cdot\mathbf{P}$. 
The solution of the above equation can be written as 
\begin{equation}\label{eq:phifnofp}
\phi_{\mathbf{r}} = \int G_{\mathbf{r},\mathbf{r'}}\left( \rho_{\mathbf{r'}} - \nabla\cdot\mathbf{P}_{\mathbf{r'}} \right)d^{3}r',
\end{equation}
where $G(\mathbf{r},\mathbf{r'})$ is the Green's function in free space which satisfies the equation:
\begin{equation}\label{eq:deltafn}
\nabla_{\mathbf{r}} ^{2}G_{\mathbf{r},\mathbf{r'}} = - 4\pi\delta(\mathbf{r}-\mathbf{r'}),
\end{equation}
and is given by
\begin{equation}\label{eq:greensfn}
G_{\mathbf{r},\mathbf{r'}} = \frac{1}{|\mathbf{r}-\mathbf{r'}|}.
\end{equation}
Note that $G(\mathbf{r},\mathbf{r'})$ also obeys DBC.
Substituting $\phi$ from \eqref{eq:phifnofp} in \eqref{eq:ve}, we obtain $\mathbf{E}$ in terms of $\mathbf{P}$:
\begin{equation}\label{eq:efnofp}
\mathbf{E}_{\mathbf{r}} = -\nabla\int G_{\mathbf{r},\mathbf{r'}}\left( \rho_{\mathbf{r'}} - \nabla\cdot\mathbf{P}_{\mathbf{r'}} \right)d^{3}r'.
\end{equation}
Using \eqref{eq:phifnofp} and \eqref{eq:efnofp} we eliminate $\phi$ and $\mathbf{E}$ from \eqref{eq:inP} to obtain a functional with $\mathbf{P}$ as the sole 
variational field:
\begin{equation}\label{eq:fnalP}
\begin{split}
\mathscr{F}\left[\mathbf{P}\right]=\int\frac{|\mathbf{P}_{\mathbf{r}}|^{2}}{2\chi_{\mathbf{r}}}d^{3}r
&+ \frac{1}{2}\iint\left( \rho_{\mathbf{r}} - \nabla\cdot\mathbf{P}_{\mathbf{r}} \right) G_{\mathbf{r},\mathbf{r'}} \\
&\times\left(\rho_{\mathbf{r'}} - \nabla\cdot\mathbf{P}_{\mathbf{r'}} \right) d^{3}r'd^{3}r.
\end{split}
\end{equation}

It can be shown that the correct constitutive relation between the polarization field and the electric field is obtained as a result of the extremization of the above functional \cite{attard}.
Furthermore, one can prove that $\mathscr{F}\left[\mathbf{P}\right]$ is an energy functional; that is, its minimum computes
the equilibrium electrostatic energy \cite{attard}.
The functional in \eqref{eq:fnalP} has been obtained previously \cite{marcus,felderhof}, but with different derivations than ours.
We now show how to transform $\mathscr{F}\left[\mathbf{P}\right]$ to an energy functional of the polarization charge density $\omega$. 
This transition begins by inserting the definition of $\omega$, namely,
\begin{equation}\label{eq:omega}
\omega_{\mathbf{r}} = - \nabla\cdot\mathbf{P}_{\mathbf{r}},
\end{equation}
in \eqref{eq:fnalP} by means of a Lagrange multiplier $\psi$:
\begin{equation}\begin{split}\label{eq:inomega}
\mathscr{F}[&\mathbf{P},\omega,\psi]=\int\frac{|\mathbf{P}_{\mathbf{r}}|^{2}}{2\chi_{\mathbf{r}}}d^{3}r
+ \frac{1}{2}\iint\left( \rho_{\mathbf{r}} + \omega_{\mathbf{r}} \right) G_{\mathbf{r},\mathbf{r'}}\\
&\times\left(\rho_{\mathbf{r'}} + \omega_{\mathbf{r'}} \right) d^{3}r'd^{3}r
-\int\psi_{\mathbf{r}}\left( \omega_{\mathbf{r}} + \nabla\cdot\mathbf{P}_{\mathbf{r}} \right) d^{3}r.
\end{split}\end{equation}
We note that $\psi$ will soon be shown to coincide with the electrostatic potential $\phi$ at equilibrium.
Taking variations of the above functional with respect to $\omega$ and $\mathbf{P}$ gives the following relations:
\begin{align}
\begin{split}\label{eq:varomega}
&\delta\omega:\qquad \psi_{\mathbf{r}} = \int G_{\mathbf{r},\mathbf{r'}} \left(\rho_{\mathbf{r'}} + \omega_{\mathbf{r'}}\right) d^{3}r', 
\end{split}
\\
\begin{split}\label{eq:varP}
&\delta\mathbf{P}:\qquad \mathbf{P}_{\mathbf{r}} = -\chi_{\mathbf{r}}\nabla\psi_{\mathbf{r}}.
\end{split}
\end{align}
Equation \eqref{eq:varomega} expresses $\psi$ in terms of $\omega$.
Substituting $\psi$ from \eqref{eq:varomega} in \eqref{eq:varP} expresses $\mathbf{P}$ in terms of $\omega$:
\begin{equation}\label{eq:Pinomega}
\mathbf{P}_{\mathbf{r}} = - \chi_{\mathbf{r}}\nabla\int G_{\mathbf{r},\mathbf{r'}}
\left( \rho_{\mathbf{r'}} + \omega_{\mathbf{r'}} \right) d^{3}r'.
\end{equation}
At this point, by using \eqref{eq:varomega} and \eqref{eq:Pinomega}, we can eliminate $\psi$ and $\mathbf{P}$ from \eqref{eq:inomega} in favor of $\omega$  
and complete the desired transformation. However, while the functional that results from this procedure does single out the correct 
physical quantity upon extremization, it becomes a maximum, not a minimum, at equilibrium. 
We elaborate more on this observation in Sec.~\ref{sec:dis}.

To obtain the functional of $\omega$ with the desired extremal behavior, one must resist substitution at this stage
and instead take the unutilized variation of $\mathscr{F}[\mathbf{P},\omega,\psi]$ with respect to $\psi$ which leads to 
\begin{equation}\label{eq:varpsi1}
\omega_{\mathbf{r}} = -\nabla\cdot \mathbf{P}_{\mathbf{r}}.
\end{equation}
Substituting $\mathbf{P}$ from \eqref{eq:Pinomega} in the above equation we obtain
\begin{equation}\label{eq:varpsi}
\omega_{\mathbf{r}} = \nabla\cdot\left\{\chi_{\mathbf{r}}\nabla\int G_{\mathbf{r},\mathbf{r'}}\left(\rho_{\mathbf{r'}} + \omega_{\mathbf{r'}}\right)d^{3}r'\right\}.
\end{equation}
The above relation, as one can tell by inspection, is the correct physical relation that the polarization charge density must satisfy.
Equations \eqref{eq:varomega} and \eqref{eq:varpsi1}, when viewed together, and compared with \eqref{eq:phifnofp} imply that $\psi$ is indeed the electrostatic potential at equilibrium. 
At this point the substitution of $\psi$, $\mathbf{P}$, and $\omega$ from \eqref{eq:varomega}, \eqref{eq:Pinomega},
and \eqref{eq:varpsi} respectively, into the functional of \eqref{eq:inomega} leads to
our central result: the functional in Eq.~\eqref{eq:fnal}.

In Appendix \ref{sec:ext} we prove that $\mathscr{F}[\omega]$ is an energy functional and its
minimization provides the correct induced charge density. 
We now analyze in some depth why some substitutions lead to an energy functional and 
why others do not.

\subsection{Key aspects of the variational principle}\label{sec:dis}
In this section we elaborate on some key observations made during the process of deriving 
$\mathscr{F}[\omega]$.
We noted in Sec.~\ref{sec:deriv} that not all substitutions to eliminate field variables from \eqref{eq:inomega} in favor of $\omega$ lead to the desired result.
We observed that $\psi$ and $\mathbf{P}$ can be eliminated from \eqref{eq:inomega} using equations \eqref{eq:varomega} and \eqref{eq:Pinomega}, thus leading to a functional with 
$\omega$ as the sole variational field.
One can show that this process results in a functional $I[\omega]$ with the functional density:  
$\rho_{\mathbf{r}}G_{\mathbf{r},\mathbf{r'}}\left(\rho_{\mathbf{r'}}+\Omega_{\mathbf{r'}}[\omega]\right) / 2 
- \omega_{\mathbf{r}} G_{\mathbf{r},\mathbf{r'}}\left(\omega_{\mathbf{r'}} - \Omega_{\mathbf{r'}}[\omega]\right) / 2$.
Upon extremization, $I[\omega]$ singles out the correct physical quantity, but becomes a maximum at equilibrium.
In fact, $I[\omega]$ is exactly the negative of the functional in Ref.~\onlinecite{allen1}, neither of which are energy functionals.

We note that functionals $\mathscr{F}[\omega]$ and $I[\omega]$ share a common structure: 
the expression for the total electrostatic energy (the first term in either functional) 
is constrained by the correct physical relation that $\omega$ must satisfy, 
namely $\omega - \Omega[\omega] = 0$.
The one but crucial difference between these functionals is in the choice of the constraint that is 
enforced by means of Lagrange multipliers. While the constraints themselves might appear equivalent, 
their different explicit forms can endow the functional with different properties.
Previous functionals that lack some desirable properties can be understood as arising from deficient 
constraint expressions. Our current formulation provides the appropriate constraint form.

It is equally important to point out that the set of substitutions that we employed in Sec.~\ref{sec:deriv}
to arrive at the desired result are not 
the only ones that lead to an energy functional.
Due to the iterative nature of Eq.~\eqref{eq:varpsi}, different sets of substitutions leading to different energy functionals are possible.
For example, resisting substitutions post Eq.~\eqref{eq:varpsi}, and instead employing 
\eqref{eq:varpsi} to replace $\omega$ with $\Omega[\omega]$ in \eqref{eq:varomega} leads to
a new relation between $\psi$ and $\omega$.
Starting with this new relation, we can execute the same cycle of steps as before to obtain new expressions for $\mathbf{P}$ and $\omega$ in terms of $\omega$.
We thus arrive at the following set of relations:
\begin{align}
\begin{split}\label{eq:newpsi}
 \psi_{\mathbf{r}} = \int G_{\mathbf{r},\mathbf{r'}} \left(\rho_{\mathbf{r'}} + \Omega_{\mathbf{r'}}[\omega]\right) d^{3}r',
\end{split}
\\
\begin{split}\label{eq:newP}
 \mathbf{P}_{\mathbf{r}} = -\chi_{\mathbf{r}}\nabla\int G_{\mathbf{r},\mathbf{r'}} \left(\rho_{\mathbf{r'}} + \Omega_{\mathbf{r'}}[\omega]\right) d^{3}r',
\end{split}
\\
\begin{split}\label{eq:newomega}
 \omega_{\mathbf{r}} = \nabla\cdot\left( \chi_{\mathbf{r}}\nabla\int G_{\mathbf{r},\mathbf{r'}}
\left( \rho_{\mathbf{r'}} + \Omega_{\mathbf{r'}}[\omega] \right) d^{3}r'\right).
\end{split}
\end{align}
We note that at equilibrium the above obtained relations for $\psi$, $\mathbf{P}$, and $\omega$ 
coincide with the corresponding relations obtained in Sec.~\ref{sec:deriv}: equations \eqref{eq:varomega}, \eqref{eq:Pinomega}, and \eqref{eq:varpsi}. 
At this point, if we substitute $\psi$, $\mathbf{P}$, and $\omega$ from \eqref{eq:newpsi}, \eqref{eq:newP}, and \eqref{eq:newomega} respectively 
into the functional of \eqref{eq:inomega}, one can show that the 
resulting functional ($\mathscr{F}^{(2)}[\omega]$; see Eq.~\eqref{eq:fnaln} below) 
is also an energy functional.

As should be evident, the above outlined cycle of steps can be repeated many times, yielding more 
energy functionals. Specifically, we find a family of functionals $\{\mathscr{F}^{(n)}[\omega]\}$ 
with $n=1,2,3,\ldots$, where the $n^{\textrm{th}}$ member has the form
\begin{equation}\begin{split}\label{eq:fnaln}
\mathscr{F}&^{(n)}[\omega]=\frac{1}{2}\iint \rho_{\mathbf{r}}G_{\mathbf{r},\mathbf{r'}}
\left(\rho_{\mathbf{r'}}+\Omega^{(n)}_{\mathbf{r'}}\right) d^{3}r'd^{3}r\\
&- \frac{1}{2}\iint \Omega^{(n)}_{\mathbf{r}} G_{\mathbf{r},\mathbf{r'}}
\left(\Omega^{(n-1)}_{\mathbf{r'}} - \Omega^{(n)}_{\mathbf{r'}} \right) d^{3}r'd^{3}r.
\end{split}\end{equation}
In Eq.~\eqref{eq:fnaln}, $\Omega^{(n)}$ is both a function of $\mathbf{r}$ and a functional
of $\omega$, but we have suppressed the functional part of the notation for brevity. 
$\Omega^{(n)}$ is given by: 
\begin{equation}\label{eq:Omegan}
\Omega_{\mathbf{r}}^{(n)} = \nabla\cdot\left( \chi_{\mathbf{r}}\nabla\int G_{\mathbf{r},\mathbf{r'}}
\left( \rho_{\mathbf{r'}} + \Omega^{(n-1)}_{\mathbf{r'}} \right) d^{3}r'\right),
\end{equation}
with $\Omega^{(0)}_{\mathbf{r}}[\omega]$ defined as 
$\Omega^{(0)}_{\mathbf{r}}[\omega] = \int\delta(\mathbf{r} - \mathbf{r'})\omega(\mathbf{r'})d^{3}r' 
= \omega$. 
Note that, by letting $n=1$ in \eqref{eq:Omegan} we obtain 
$\Omega^{(1)}[\omega] = \Omega[\omega]$, where $\Omega[\omega]$ is given by Eq.~\eqref{eq:Omega}.

It can be shown that for every $n$ the functional given by \eqref{eq:fnaln} is an energy functional. 
Although these functionals are different from one another, upon extremization each of them 
give the same iterative relation, 
Eq.~\eqref{eq:varpsi}, and all minimize to the true electrostatic energy. 
Proofs of these assertions are similar to the ones that appear in Appendix \ref{sec:ext}.
By letting $n=1$ in \eqref{eq:fnaln} it is easy to see that 
$\mathscr{F}^{(1)}[\omega] = \mathscr{F}[\omega]$.
$\mathscr{F}[\omega]$ thus represents the simplest member of a large family of energy functionals, 
offering the most ease with regards to use in 
analytical and numerical minimization procedures. 
We will only work with $\mathscr{F}[\omega]$ in the rest of this paper. 

Each member of the family $\{\mathscr{F}^{(n)}[\omega]\}$ possesses 
the same basic structure alluded to before: 
to a term representing the electrostatic energy (the first double integral in \eqref{eq:fnaln}), 
the iterative relation that $\omega$ must satisfy is included as 
a constraint. 
As is evident from \eqref{eq:fnaln}, for each iterative relation that supplies the constraint equation, 
our variational formalism finds the appropriate Lagrange multiplier required to enforce this constraint 
such that the resulting functional acquires the desired extremal properties.

\section{Sharp dielectric interfaces}\label{sec:pud}
The functional $\mathscr{F}[\omega]$ derived in Sec.~\ref{sec:vform} works for any medium 
with linear dielectric response, even for arbitrary spatial variations.
In many instances, it is sufficient to represent the real system by a coarse-grained model where 
regions of uniform, but different, dielectric response are separated from each other by interfaces that can be assumed to be thin.
For example, in the problem of colloids in a polar solvent, modelling the colloid as 
one uniform dielectric continuum and the surrounding solvent as another uniform dielectric 
of different permittivity provides a good representation of the real system.
Other examples where coarse-graining of this kind is often employed 
include: oil-water emulsions and biopolymers, such as lipid bilayers, in aqueous solution.
In this light, we now consider the application of our functional to the problem of ions present 
in a system exhibiting this piecewise-uniform dielectric response.
We show that for this specific dielectric response the functional $\mathscr{F}[\omega]$ reduces
to a functional with only the interfacial induced charge density as the variational field.

\subsection{The functional for the case of piecewise-uniform dielectric response}
\begin{figure}[h]
\centerline{
\includegraphics[scale=0.318]{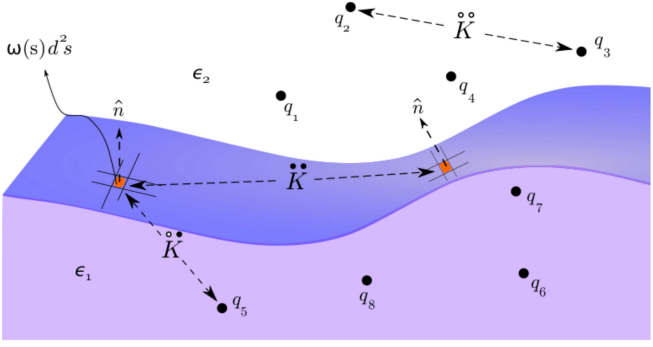}
}
\caption
{\label{fig2}
Sketch showing two uniform dielectric regions of permittivities $\epsilon_{1}$ and $\epsilon_{2}$ separated by a sharp interface that 
may assume an arbitrary shape. The figure also shows the presence of point charges in either media and the induced charge on the interface.
The effective interactions appearing in Eq.~\eqref{eq:sfnal} between the various charged constituents of the system are also shown.
}
\end{figure}

For the sake of brevity, we restrict ourselves to two uniform dielectrics separated by a single sharp interface $\mathcal{I}$, see Fig.~\ref{fig2}.
Extension to multiple dielectrics is straightforward.
We assume that ions reside in the bulk of either dielectric. Note that the interface can assume arbitrary geometry.
Let $\epsilon_{1}$ and $\epsilon_{2}$ denote the permittivities of the two media. We consider ions to be point particles with the $i^{\textrm{th}}$ ion having a charge $q_{i}$.
For a system with $N$ ions, the free charge density can be written as $\rho(\mathbf{r}) = \sum_{i=1}^{N} q_{i}\delta\left(\mathbf{r}-\mathbf{r}_{i}\right)$,
where $\mathbf{r}_{i}$ prescribes the position of the $i^{\textrm{th}}$ ion.
It is useful to define the permittivity at the interface, taken to be the mean of permittivities on either side: 
$\epsilon_{m}=(\epsilon_{1}+\epsilon_{2})/2$;
and introduce $\epsilon_{d}$ = $|\epsilon_{2}-\epsilon_{1}|/4\pi$ as a measure of the permittivity difference across $\mathcal{I}$.

Clearly, the gradient of $\epsilon$, or equivalently $\chi$, vanishes everywhere except at the interface. 
Mathematically we express this as:
\begin{equation}\label{eq:delchi}
\nabla\chi_{\mathbf{r}} = \epsilon_{d}\int_{\mathcal{I}} \hat{n}_{\mathbf{r}}\delta\left(\mathbf{r}-\mathbf{s}\right)d^{2}s,
\end{equation}
where $\hat{n}$ is the unit normal vector at the interface, chosen to point in the direction of increasing permittivity, and 
$\mathbf{s}$ is the position vector of an arbitrary point on the interface.
When $\mathbf{r}\notin\mathcal{I}$, $\nabla\chi\left(\mathbf{r}\right)=0$, otherwise $\nabla\chi\left(\mathbf{r}\right)=\epsilon_{d}\hat{n}(\mathbf{r})$.  
Because each medium offers a uniform dielectric response, the induced charge density in the bulk is known analytically 
from simple electrostatics principles \cite{jackson}, and it is:
\begin{equation}\label{eq:bulkomega} 
\omega^{\textrm{bulk}}_{\mathbf{r}} = -\frac{\epsilon_{\mathbf{r}} - 1}{\epsilon_{\mathbf{r}}}\rho_{\mathbf{r}}.
\end{equation}
It is well known that the above bulk contributon leads to an effective charge density of $\rho/\epsilon$.
Due to the discontinuity in the permittivity at the interface, induced charges also exist on the interface and their magnitude
is in general unknown. 
Thus, the overall induced charge density is expressed as the sum of two terms: 
\begin{equation}\label{eq:pudomega}
\omega_{\mathbf{r}} = \omega^{\textrm{bulk}}_{\mathbf{r}}
+\int_{\mathcal{I}} \omega_{\mathbf{r}}\delta\left(\mathbf{r}-\mathbf{s}\right)d^{2}s,
\end{equation}
where the first term on the right hand side in \eqref{eq:pudomega} is given by \eqref{eq:bulkomega}, and the second term is the interfacial (surface) induced charge density.
Similar to the mathematical representation of $\nabla\chi$, we have expressed the interfacial induced charge density as a surface integral, such that 
when $\mathbf{r}\notin \mathcal{I}$ the integral vanishes. 

Substituting $\omega$ from \eqref{eq:pudomega} in \eqref{eq:fnal},
and using Eq.~\eqref{eq:delchi}, we find that 
several volume integrals in \eqref{eq:fnal} reduce to surface integrals and 
$\mathscr{F}[\omega]$ is transformed to a functional of the surface induced charge density:
\begin{align}\label{eq:sfnal}
\mathscr{F}[\omega_{\mathbf{s}}] &= \frac{1}{2}\iint\rho_{\mathbf{r}}K^{^{^{\negthickspace\negthickspace\negmedspace\negthickspace\circ\circ}}}_{\mathbf{r},\mathbf{r'}}\rho_{\mathbf{r'}}d^{3}rd^{3}r'
+ \frac{1}{2}\iint_{\mathcal{I}}\rho_{\mathbf{r}}K^{^{^{\negthickspace\negthickspace\negmedspace\negthickspace\circ\bullet}}}_{\mathbf{r},\mathbf{s}}\omega_{\mathbf{s}}d^{3}rd^{2}s
\nonumber\\
&+ \frac{1}{2}\int_{\mathcal{I}}\int_{\mathcal{I}}\omega_{\mathbf{s}}K^{^{^{\negthickspace\negthickspace\negmedspace\negthickspace\bullet\bullet}}}_{\mathbf{s},\mathbf{s'}}\omega_{\mathbf{s'}}d^{2}sd^{2}s',
\end{align}
where $\omega(\mathbf{s})$ is the induced charge density at the position $\mathbf{s}$ on the interface, and
$K^{^{^{\negthickspace\negthickspace\negmedspace\negthickspace\circ\circ}}}$, $K^{^{^{\negthickspace\negthickspace\negmedspace\negthickspace\circ\bullet}}}$, 
and $K^{^{^{\negthickspace\negthickspace\negmedspace\negthickspace\bullet\bullet}}}$ are, respectively, 
the effective potentials of interaction between two free charges, between
a free charge and an induced charge, and between two induced charges (see Fig.~\ref{fig2}). 
These effective interactions are given by:
\begin{align}\label{eq:K}
&K^{^{^{\negthickspace\negthickspace\negmedspace\negthickspace\circ\circ}}}_{\mathbf{r},\mathbf{r'}} = \frac{1}{\epsilon_{\mathbf{r}}}G_{\mathbf{r},\mathbf{r'}}
\,+\, \frac{1}{\epsilon_{\mathbf{r}}}\,\overline{G}_{\mathbf{r},\mathbf{r'}}\,\frac{1}{\epsilon_{\mathbf{r'}}}
\,+\, \frac{1}{\epsilon_{\mathbf{r}}}\,\overline{\overline{G}}_{\mathbf{r},\mathbf{r'}}\,\frac{1}{\epsilon_{\mathbf{r'}}}
\nonumber\\
&K^{^{^{\negthickspace\negthickspace\negmedspace\negthickspace\circ\bullet}}}_{\mathbf{r},\mathbf{s}} = \frac{\epsilon_{\mathbf{r}} - \epsilon_{m}}{\epsilon_{\mathbf{r}}} G_{\mathbf{r},\mathbf{s}}
+ \frac{\overline{G}_{\mathbf{s},\mathbf{r}} - \left(2\epsilon_{m}-1\right)\overline{G}_{\mathbf{r},\mathbf{s}}}{\epsilon_{\mathbf{r}}}
+ \frac{2\overline{\overline{G}}_{\mathbf{r},\mathbf{s}}}{\epsilon_{\mathbf{r}}}
\nonumber\\
&K^{^{^{\negthickspace\negthickspace\negmedspace\negthickspace\bullet\bullet}}}_{\mathbf{s},\mathbf{s'}} = \epsilon_{m}\left(\epsilon_{m}-1\right)G_{\mathbf{s},\mathbf{s'}} -
\left(2\epsilon_{m}-1\right)\overline{G}_{\mathbf{s},\mathbf{s'}} +
\overline{\overline{G}}_{\mathbf{s},\mathbf{s'}}.
\end{align}
While the function $G$ in \eqref{eq:K} is the bare Green's function given by \eqref{eq:greensfn}, we find two new potentials of interaction in \eqref{eq:K}, 
$\overline{G}$ and $\overline{\overline{G}}$, which are defined as:
\begin{align}\label{eq:modG}
&\overline{G}_{\mathbf{a},\mathbf{b}} = \epsilon_{d}\int_{\mathcal{I}} G_{\mathbf{a},\mathbf{u}} \; \hat{n}_{\mathbf{u}}\cdot\nabla_{\mathbf{u}} G_{\mathbf{u},\mathbf{b}} \; d^{2}u
\nonumber\\
&\overline{\overline{G}}_{\mathbf{a},\mathbf{b}}=
\epsilon_{d}^{2}\int_{\mathcal{I}}\int_{\mathcal{I}}\hat{n}_{\mathbf{u}}\cdot\nabla_{\mathbf{u}} G_{\mathbf{a},\mathbf{u}} \,G_{\mathbf{u},\mathbf{v}}\, 
\hat{n}_{\mathbf{v}}\cdot\nabla_{\mathbf{v}} G_{\mathbf{v},\mathbf{b}} \,d^{2}ud^{2}v,
\end{align}
where $\mathbf{a}, \mathbf{b}$ are arbitrary position vectors and
$\mathbf{u}, \mathbf{v}$ are position vectors of arbitrary interfacial points.

The functional in Eq.~\eqref{eq:sfnal} can be compared with the functional of 
the surface polarization charge density obtained in Ref.~\onlinecite{allen1}. 
The latter functional, as we noted earlier in Sec.~\ref{sec:dis}, is not an energy functional. 
We find that the major difference between these two functionals is the \emph{absence} of the
interaction $\overline{\overline{G}}$ in the functional of Ref.~\onlinecite{allen1}. 
We note that the presence of this particular interaction potential in $\mathscr{F}[\omega_{\mathbf{s}}]$
is the direct result of employing the appropriate choice for the constraint to the electrostatic energy, an 
aspect of the variational formulation we highlighted in Sec.~\ref{sec:dis}.
It also appears that a functional of polarization charge density 
constructed with a combination of only $G$ and $\overline{G}$ interaction
potentials (like the one in Ref.~\onlinecite{allen1}) ceases to remain an energy functional; although
we have not been able to rigorously show this. 
Our attempts to construct an energy functional involving 
only $G$ and $\overline{G}$ interactions via the variational formulation presented here, or otherwise,
failed.

We now employ $\mathscr{F}[\omega_{\mathbf{s}}]$ to study some simple 
model systems exhibiting piecewise-uniform dielectric response. 
For these systems we can analytically carry out the integrals 
involved in Eq.~\ref{eq:sfnal} by finding a suitable basis and 
expanding the Green's function and induced charge density in terms of the associated basis functions;
eventually expressing the functional as a single integral involving the undetermined coefficients
of expansion.
We show that for these solvable models, $\mathscr{F}[\omega_{\mathbf{s}}]$ can be minimized analytically 
which leads to the well known expressions for the associated induced charge density.

\subsection{A point charge near a thin planar wall}
\begin{figure}[h]
\centerline{
\includegraphics[scale=0.4]{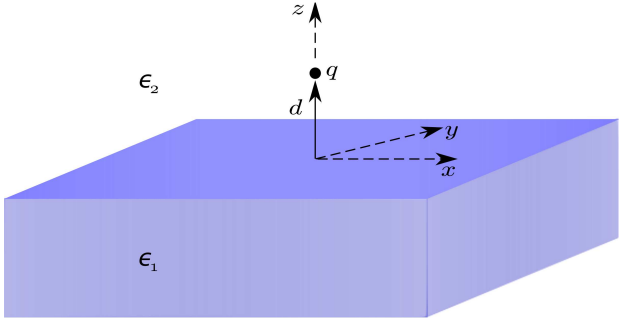}
}
\caption
{\label{fig3}
Sketch showing a flat interface (the $z=0$ plane) separating the region $z<0$ of permittivity $\epsilon_{1}$ with the region $z>0$
characterized by a different dielectric constant $\epsilon_{2}$. 
A point charge $q$ is present near the interface at a distance $d$ from the origin in the medium 
with permittivity $\epsilon_{2}$.
}
\end{figure}

We consider a planar interface separating two dielectrics of 
different permittivities $\epsilon_{1}$ and $\epsilon_{2}$, with $\epsilon_{1}>\epsilon_{2}$ assumed. 
A point particle of charge $q$ is placed at a distance $d$ from the interface in the region with lower permittivity (see Fig.~\ref{fig3}). 
We derive the induced charge density at the interface for this system using our variational formalism,
in the process revealing the expression for the functional for this specific case. 

We adopt cylindrical coordinates $(\mathcal{R},\phi,z)$, assume the interface to be the $z=0$ plane, 
and take the point charge to be on the positive $z$ axis at a distance $d$ from the origin. 
The $z>0$ domain then becomes the dielectric with $\epsilon_{2}$ permittivity.
It is useful to choose Bessel functions of integer order as the basis functions in this case.
In this basis the bare Green's function can be expanded as:
\begin{equation}\label{eq:gplanar}
G_{\mathbf{r},\mathbf{r'}} = 
\sum_{m=-\infty}^{\infty} e^{im(\phi - \phi')}\int_{0}^{\infty}J_{m}(\alpha\mathcal{R})J_{m}(\alpha\mathcal{R}')e^{-\alpha\left|z - z'\right|}d\alpha.
\end{equation}
We note that the position vector of the point charge is $\mathbf{r}_{q} = (0,0,d)$ and the position vector of 
a point on the interface has the form $\mathbf{s}=(\mathcal{R},\phi,0)$. Also, given the set up of this problem and recalling
the definition of the normal vector (it points from lower permittivity dielectric to the higher one), we have $\hat{n} = - \hat{z}$.
Let us now evaluate the functional given in \eqref{eq:sfnal} for this particular different dielectric problem. 
For this purpose we would need the Green's functions
$G(\mathbf{r}_{q},\mathbf{s})$ and $G(\mathbf{s},\mathbf{s'})$, and the dot product of their gradients with the normal vector $\hat{n}$. 
These are readily evaluated from \eqref{eq:gplanar}
by employing, wherever necessary, the properties of Bessel functions: $J_{0}(0)=1$, $J_{m\neq0}(0)=0$.
Using these functions the necessary
renormalized Green's functions $\overline{G}$ and $\overline{\overline{G}}$ can be evaluated from \eqref{eq:modG}, and employing them
in \eqref{eq:K}, the effective interactions 
$K^{^{^{\negthickspace\negthickspace\negmedspace\negthickspace\circ\circ}}}$, $K^{^{^{\negthickspace\negthickspace\negmedspace\negthickspace\circ\bullet}}}$, 
and $K^{^{^{\negthickspace\negthickspace\negmedspace\negthickspace\bullet\bullet}}}$ are known.
Finally, just like the Green's function in \eqref{eq:gplanar}, the induced charge density, which due to symmetry is only a function of
$\mathcal{R}$, can be written as an integral involving Bessel function $J_{0}$:
\begin{equation}\label{eq:wexpansion}
w(\mathcal{R}) = \int_{0}^{\infty}A(\alpha)J_{0}(\alpha\mathcal{R})d\alpha,
\end{equation}
where $A(\alpha)$ are as of now undetermined.
Submitting the evaluated interactions
$K^{^{^{\negthickspace\negthickspace\negmedspace\negthickspace\circ\circ}}}$, $K^{^{^{\negthickspace\negthickspace\negmedspace\negthickspace\circ\bullet}}}$, 
and $K^{^{^{\negthickspace\negthickspace\negmedspace\negthickspace\bullet\bullet}}}$, 
and $\omega(\mathbf{s})\equiv\omega(\mathcal{R})$ 
from the above equation in the functional of \eqref{eq:sfnal}, and remembering that the area element in our chosen coordinates is $\mathcal{R}d\mathcal{R}d\phi$,
we carry out most of the resulting integrals by employing the orthogonality relation
\begin{equation}
\int_{0}^{\infty} J_{m}(\alpha\mathcal{R})J_{m}(\alpha\mathcal{R'})d\alpha = \frac{1}{\mathcal{R}}\delta(\mathcal{R}-\mathcal{R'}),
\end{equation}
and obtain the following functional:
\begin{equation}\begin{split}\label{eq:planarf}
\mathscr{F}_{|}[A&(\alpha)]=
\frac{q^{2}}{8\epsilon_{2}^{2}}\left(\epsilon_{1}-\epsilon_{2}\right)\left(\epsilon_{1}-\epsilon_{2}-2\right)\int_{0}^{\infty}e^{-2\alpha d}d\alpha\\
&+\frac{q\pi}{2\epsilon_{2}}\left(\epsilon_{1}-\epsilon_{2}\right)\left(\epsilon_{1}+\epsilon_{2}-2\right)\int_{0}^{\infty}\frac{A\left(\alpha\right)}{\alpha}e^{-\alpha d}d\alpha\\ 
&+\frac{\pi^{2}}{2}\left(\epsilon_{1}+\epsilon_{2}\right)\left(\epsilon_{1}+\epsilon_{2}-2\right)\int_{0}^{\infty}\frac{A^{2}\left(\alpha\right)}{\alpha^{2}}d\alpha.
\end{split}\end{equation}
The subscript on $\mathscr{F}$ in \eqref{eq:planarf} represents that the above is a functional for the planar interface case.

Clearly, $\mathscr{F}_{|}[A(\alpha)]$ is a functional of the lone function variable $A(\alpha)$, which
through \eqref{eq:wexpansion} represents the induced surface charge density.
We now take the functional derivative of $\mathscr{F}_{|}[A(\alpha)]$ and set it to zero in order to determine $A(\alpha)$. 
It is obvious that only the last two terms in \eqref{eq:planarf} contribute to this process and we obtain 
\begin{equation}\begin{split}\label{eq:Diplanar}
\frac{\delta\mathscr{F}_{|}[A(\alpha)]}{\delta A(\alpha)}&=
\frac{q\pi}{2\epsilon_{2}}\left(\epsilon_{1}-\epsilon_{2}\right)\left(\epsilon_{1}+\epsilon_{2}-2\right)\frac{e^{-\alpha d}}{\alpha}\\
&+\pi^{2}\left(\epsilon_{1}+\epsilon_{2}\right)\left(\epsilon_{1}+\epsilon_{2}-2\right)\frac{A\left(\alpha\right)}{\alpha^{2}},
\end{split}\end{equation}
which when set to zero reveals the coefficients $A(\alpha)$ to be the following:
\begin{equation}\label{eq:Aplanar}
A(\alpha) = -\frac{\epsilon_{1}-\epsilon_{2}}{\epsilon_{1}+\epsilon_{2}}\frac{q}{2\pi\epsilon_{2}}\alpha e^{-\alpha d}.
\end{equation}
Using the expression for $A(\alpha)$ from above in \eqref{eq:wexpansion} and carrying out the single integral involving the zeroth order Bessel function,
we get 
\begin{equation}\label{eq:wplanar}
\omega(\mathbf{s})\equiv\omega(\mathcal{R}) = -\frac{\epsilon_{1}-\epsilon_{2}}{\epsilon_{1}+\epsilon_{2}}\frac{q}{2\pi\epsilon_{2}}\frac{d}{\left(\mathcal{R}^{2}+d^{2}\right)^{3/2}}.
\end{equation}
The above expression matches with the standard result for the surface induced charge density (see Ref.~\onlinecite{jackson}). 

\subsection{A point charge near a thin spherical interface}
We now derive the functional form and induced density for the case of a point charge near a spherical dielectric. 
The derivation is similar to the one just carried out for the planar case.
 
\begin{figure}[h]
\centerline{
\includegraphics[scale=0.4]{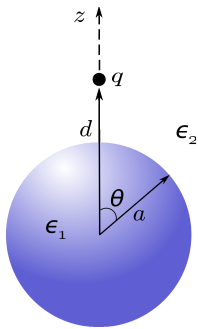}
}
\caption
{\label{fig4}
Sketch of a dielectric sphere of radius $a$ and permittivity $\epsilon_{1}$ embedded
in a region characterized by dielectric permittivity $\epsilon_{2}$. The sphere
is centered at the origin and is oriented such that the $z$ axis coincides with $\theta=0$, where
$\theta$ is the polar angle. A point charge $q$ is placed outside the dielectric sphere, on the $z$-axis at a distance $d$ from 
the origin. 
}
\end{figure}

We consider a dielectric sphere of radius $a$ and permittivity $\epsilon_{1}$ 
surrounded by a different dielectric of permittivity $\epsilon_{2}$ (where $\epsilon_{1} > \epsilon_{2}$ is assumed). 
There is a free charge $q$ placed outside the sphere. 
In this example, the use of spherical coordinates $r, \theta, \phi$ is most convenient. 
We take the sphere to center at the origin $r = 0$ and the point charge is assumed to lie on the positive $z$ axis at a distance $d$ from the origin (see Fig.~\ref{fig4}), giving it 
the position vector $\mathbf{r}_{q}=(d, 0, 0)$. Our conventions result in the the unit normal vector
to be $\hat{n} = -\hat{r}$, i.e, pointing into the sphere.

We start by choosing a suitable basis to expand the Green's function.
This basis turns out to be spherical harmonics and 
the expansion of $G(\mathbf{r},\mathbf{r'})$ in the latter is given by:
\begin{equation}\label{eq:gsphere}
G(\mathbf{r},\mathbf{r'}) = 4\pi\sum_{l=0}^{\infty}\frac{1}{2l+1}\frac{r^{l}}{r'^{l+1}}\sum_{m=-l}^{l}Y_{lm}(\theta,\phi)Y_{lm}^{*}(\theta',\phi'),
\end{equation}
where $r=|\mathbf{r}|$, $r'=|\mathbf{r'}|$, and the above expression holds for $r \le r'$.
As dictated by equation \eqref{eq:modG}, 
in order to evaluate the interactions $\overline{G}$ and $\overline{\overline{G}}$ for this problem, we need 
the following Green's functions and its derivatives: $G(\mathbf{s},\mathbf{r}_{q})$, $\hat{n}\cdot\nabla G(\mathbf{s},\mathbf{r}_{q})$, 
$G(\mathbf{s},\mathbf{s'})$, and $\hat{n}\cdot\nabla G(\mathbf{s},\mathbf{s'})$, where $\mathbf{s}$ and $\mathbf{s'}$ are the position 
vectors of arbitrary points on the sphere.
These functions are readily evaluated from equation \eqref{eq:gsphere} by employing the relations: 
$Y_{lm}(0,0) = 0$ for $m\neq0$, and $Y_{l0}(0,0) = \sqrt\frac{2l+1}{4\pi}$.
Once $\overline{G}$ and $\overline{\overline{G}}$ are known, the effective interactions 
$K^{^{^{\negthickspace\negthickspace\negmedspace\negthickspace\circ\circ}}}$, $K^{^{^{\negthickspace\negthickspace\negmedspace\negthickspace\circ\bullet}}}$, 
and $K^{^{^{\negthickspace\negthickspace\negmedspace\negthickspace\bullet\bullet}}}$
for this problem are computed from \eqref{eq:K}.
These interactions are then plugged into the functional in \eqref{eq:sfnal}.
Also, for symmetry reasons the surface induced charge density $\omega(\mathbf{s})$ is independent of the variable
$\phi$ and just like any regular function of $\theta$, it can be expanded in terms of spherical harmonics as:
\begin{equation}\label{eq:wsphere1}
\omega(\theta) = \sum_{l=0}^{\infty}A_{l}P_{l}(\cos\theta) = \sum_{l=0}^{\infty}\sqrt{\frac{4\pi}{2l+1}}A_{l}Y_{l0}(\theta),
\end{equation}
where the coefficients of expansion $A_{l}$ are as yet unknown. Using \eqref{eq:wsphere1} in \eqref{eq:sfnal}, and employing the orthonormality relation
\begin{equation}
\int_{0}^{2\pi}\int_{0}^{\pi} Y_{lm}(\theta,\phi)Y_{l'm'}^{*}(\theta,\phi)\sin\theta d\theta d\phi = \delta_{mm'}\delta_{ll'}
\end{equation}
wherever necessary, the functional for this particular different dielectric problem is found to be:
\begin{equation}\begin{split}\label{eq:fsphere}
\mathscr{F}_{\circ}[\{A_{l}\}]
&=\frac{q^{2}}{2\epsilon_{2}^{2}}\sum_{l=0}^{\infty}\frac{\epsilon_{1}-\epsilon_{2}}{2l+1}l
\left(\frac{\epsilon_{1}-\epsilon_{2}}{2l+1}l - 1\right)\frac{a^{2l+1}}{d^{2l+2}}\\
&+\frac{q}{2\epsilon_{2}}\sum_{l=0}^{\infty}\frac{4\pi}{2l+1}\frac{\epsilon_{1}-\epsilon_{2}}{2l+1}l(b-2)\frac{a^{l+2}}{d^{l+1}}A_{l}\\
&+\frac{a^{3}}{8}\sum_{l=0}^{\infty}\left(\frac{4\pi}{2l+1}\right)^{2}b(b-2)A^{2}_{l},
\end{split}\end{equation}
where 
\begin{equation}\label{eq:bsphere}
b = \epsilon_{1} + \epsilon_{2} - \frac{\epsilon_{1}-\epsilon_{2}}{2l+1},
\end{equation}
and the subscript on $\mathscr{F}$ in \eqref{eq:fsphere} denotes the spherical interface case under study.

Evaluating the functional derivative of $\mathscr{F}_{\circ}[\{A_{l}\}]$ and 
setting it to zero leads up to an equation for $A_{l}$, which upon 
subsequent solving for, gives
\begin{equation}
A_{l} = - \frac{2q}{\epsilon_{2}}\frac{\epsilon_{1}-\epsilon_{2}}{4\pi b}l\frac{a^{l-1}}{d^{l+1}}.
\end{equation}
Substituting $b$ from \eqref{eq:bsphere} in the above equation leads to
\begin{equation}\label{eq:Al}
A_{l} = - \frac{q}{4\pi\epsilon_{2}}\frac{(\epsilon_{1}-\epsilon_{2})l(2l+1)}{l(\epsilon_{1}+\epsilon_{2}) + \epsilon_{2}}\frac{a^{l-1}}{d^{l+1}}.
\end{equation}
Plugging $A_{l}$ from \eqref{eq:Al} in \eqref{eq:wsphere1} gives the induced charge density on the interface to be
\begin{equation}\label{eq:wsphere2}
\omega(\theta) = -\frac{q}{4\pi\epsilon_{2}}\sum_{l=0}^{\infty}\frac{(\epsilon_{1}-\epsilon_{2})l(2l+1)}{l(\epsilon_{1}+\epsilon_{2}) + \epsilon_{2}}\frac{a^{l-1}}{d^{l+1}}P_{l}(\cos\theta),
\end{equation}
which matches with the standard result available elsewhere \cite{allen1}.

\section{Numerical minimization of $\mathscr{F}[\omega_{\mathbf{s}}]$}\label{sec:numeric}
When many point charges are present near an arbitrarily shaped dielectric interface,
one must resort to numerical methods to minimize $\mathscr{F}[\omega_{\mathbf{s}}]$ in order to compute
the induced charge density on the interface. 
Therefore, we now turn towards discussing the numerical implementation of our variational method. 

To perform the minimization numerically we first partition the dielectric interface into $M$ finite elements. 
To each element $k$ we assign an average induced charge density $\omega_{k}$, an area $a_{k}$ and a normal vector $n_{k}$.
Under this discrete representation, $\mathscr{F}[\omega_{\mathbf{s}}]$ becomes a functional 
of the set of discrete induced charge density values $\{\omega_{k}\}$ and Eq. \eqref{eq:sfnal} is transformed into:
\begin{align}\label{eq:dsfnal}
\mathscr{F}[\{\omega_{k}\}] &= \frac{1}{2}\sum_{i=1}^{N}\sum_{j=1\atop j\neq i}^{N}q_{i}
K^{^{^{\negthickspace\negthickspace\negmedspace\negthickspace\circ\circ}}}_{\mathbf{r}_{i},\mathbf{r}_{j}}q_{j}
+ \frac{1}{2}\sum_{i=1}^{N}\sum_{k=1}^{M}q_{i}
K^{^{^{\negthickspace\negthickspace\negmedspace\negthickspace\circ\bullet}}}_{\mathbf{r}_{i},\mathbf{s}_{k}}\omega_{k}a_{k}
\nonumber\\
&+ \frac{1}{2}\sum_{k=1}^{M}\sum_{l=1}^{M}\omega_{k}
K^{^{^{\negthickspace\negthickspace\negmedspace\negthickspace\bullet\bullet}}}_{\mathbf{s}_{k},\mathbf{s}_{l}}
\omega_{l}a_{k}a_{l},
\end{align}
where $\mathbf{s}_{k}$ is the position vector of the $k^{\textrm{th}}$ finite element and $N$ is the number of 
point charges. 
Note that we represent the point charges with the density 
$\rho(\mathbf{r})=\sum_{i=1}^{N}q_{i}\delta(\mathbf{r} - \mathbf{r}_{i})$, where $q_{i}$ and $\mathbf{r}_{i}$ 
are, respectively, the charge and position vector of the $i^{\textrm{th}}$ point charge. 
The effective interactions
$K^{^{^{\negthickspace\negthickspace\negmedspace\negthickspace\circ\circ}}}$, $K^{^{^{\negthickspace\negthickspace\negmedspace\negthickspace\circ\bullet}}}$, 
and $K^{^{^{\negthickspace\negthickspace\negmedspace\negthickspace\bullet\bullet}}}$ in the above equation are
the discretized version of their continuum counterparts in Eq. \eqref{eq:K}.
We note that the discretization process introduces divergences in \eqref{eq:dsfnal}, for example when $k=l$, and 
to get around these divergences, we replace the sum at these points by an approximate integral which is 
evaluated analytically \cite{allen1}.

$\mathscr{F}[\{\omega_{k}\}]$ can now be minimized by using steepest descent or simulated annealing methods.
We choose the procedure of simulated annealing and implement it using a molecular dynamics (MD) scheme \cite{madden}. 
We include $\mathscr{F}[\{\omega_{k}\}]$ as the potential energy part of a Lagrangian that contains
a fictitious kinetic energy term: $\sum_{k=1}^{M}\mu_{k}\dot{\omega}_{k}^{2}/2$, where $\mu_{k}$ is a fictitious mass 
assigned to the $k^{\textrm{th}}$ induced charge value. 
The set $\{\omega_{k}\}$ represents a point in the (fictitious) configuration space and 
equations of motion of this point are derived from the Lagrangian. 
A feature of the system that becomes important in simulations is that,
as a result of Gauss's law, the net induced charge at the interface is a constant. 
We directly enforce this constraint at each step of the simulation via the shake-rattle algorithm \cite{shake}.
\begin{figure}[h]
\centerline{
\includegraphics[scale=0.4]{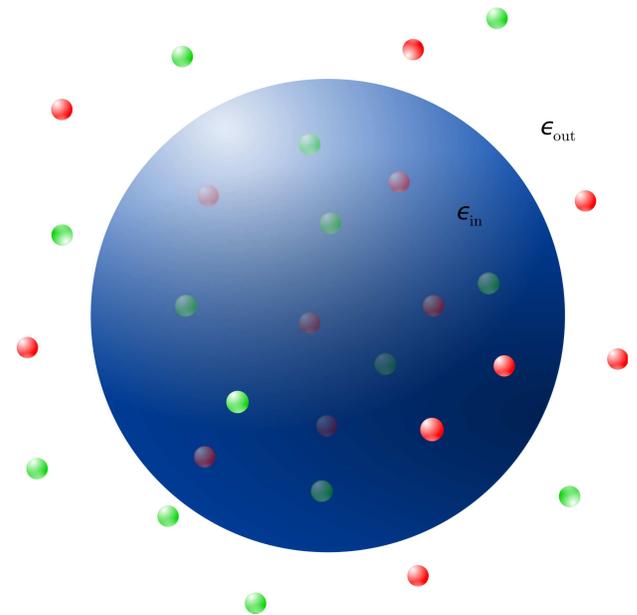}
}
\caption
{\label{fig6}
The system for testing the numerical procedure designed to minimize $\mathscr{F}[\omega_{\mathbf{s}}]$: 
positive (red) and negative (green) ions inside and outside a spherically shaped dielectric
region. 
The dielectric permittivity inside (outside) the sphere is $\epsilon_{\textrm{in}}$ ($\epsilon_{\textrm{out}}$). 
The ions have the charge of $\pm 1e$ and are represented as spheres of diameter $\sigma$; the radius of the dielectric sphere
is $a=10\sigma$. See text for the meaning of the symbols.
}
\end{figure}
\begin{figure}
\centerline{
\includegraphics[scale=0.6]{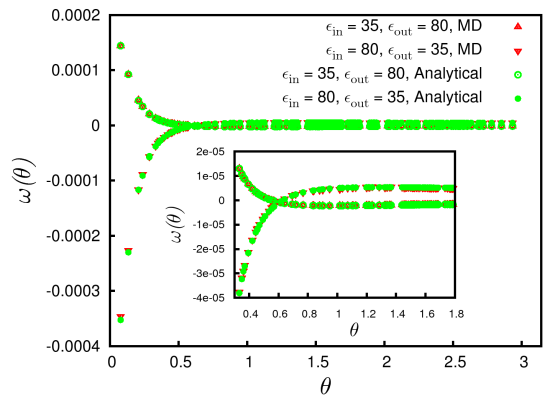}
}
\caption
{\label{fig5}
Polarization charge density (in units of $e/\sigma^{2}$) as a function of $\theta$ 
induced on the dielectric sphere from a point charge located outside the sphere (see Fig.~\ref{fig4}).
$\epsilon_{\textrm{in}}$ is the permittivity inside the sphere and 
$\epsilon_{\textrm{out}}$ is the permittivity outside. 
Red triangles are the results from the numerical minimization of the 
functional $\mathscr{F}[\omega_{\mathbf{s}}]$ and 
green circles are exact results given by Eq.~\eqref{eq:wsphere2}. 
The standard triangles and the hollow circles correspond
to the case when $\epsilon_{\textrm{in}} = 35$ and $\epsilon_{\textrm{out}} = 80$, 
while the inverted triangles and filled 
circles correspond to the inverse problem where 
$\epsilon_{\textrm{in}} = 80$ and $\epsilon_{\textrm{out}} = 35$.
The inset shows the details of the density around the point where it changes sign.   
}
\end{figure}

The simulation begins at an arbitrarily chosen point in the fictitious configuration space and we choose 
$\mu_{k}$ and the simulation time step $\Delta$ such that the ensuing dynamics is stable. 
The dynamics of this point is generated via standard MD algorithm, using the force obtained 
as a result of computing the gradient of the functional in \eqref{eq:dsfnal} with respect to $\omega_{k}$. 
The motion of the point is towards the minimum of the potential energy, 
resulting in the rise of the fictitious kinetic energy. 
After some time, a fraction of the kinetic energy is removed from the system, 
and the whole process of exploring the configuration space begins again. 
Eventually, the system reaches its minimum potential energy 
and the set of induced charge values corresponding to this state is obtained as the solution.

% new
To demonstrate and test our numerical optimization strategy, we first applied it to the problem of a single positive 
charge outside a dielectric sphere as depicted in Fig.~\ref{fig4}, and then to the problem of many charges near
a spherical dielectric interface as shown in Fig.~\ref{fig6}. 
For the single test charge problem, the exact result for the induced density is given by Eq.~\eqref{eq:wsphere2}.
The exact results for the induced charge density in the many-charge system are obtained from a careful 
superposition of the induced densities generated by considering each charge separately. 
We test the accuracy of our numerical procedure against these exact results.
We consider a spherical dielectric of permittivity $\epsilon_{\textrm{in}}$ surrounded by an exterior dielectric with 
permittivity $\epsilon_{\textrm{out}}$. The radius of the sphere is $a=10\sigma$, where $\sigma$ is the diameter
of the point charge taken to be $\sigma=l_{B}/2$ and serves as the length unit. Here, $\l_{B}$ is the Bjerrum length in 
water. The unit of charge is taken to be $e$, the charge on a proton. 
The interface is discretized with roughly $M=600$ points and the fictitious MD simulation parameters are: 
$\Delta = 0.001$, $\mu_{k} = 5 - 10$, and $S=100,000$. 
$S$ is the number of MD steps and we quench the system every $S/10$ steps.

\begin{figure*}
\centerline{
\includegraphics[scale=0.6]{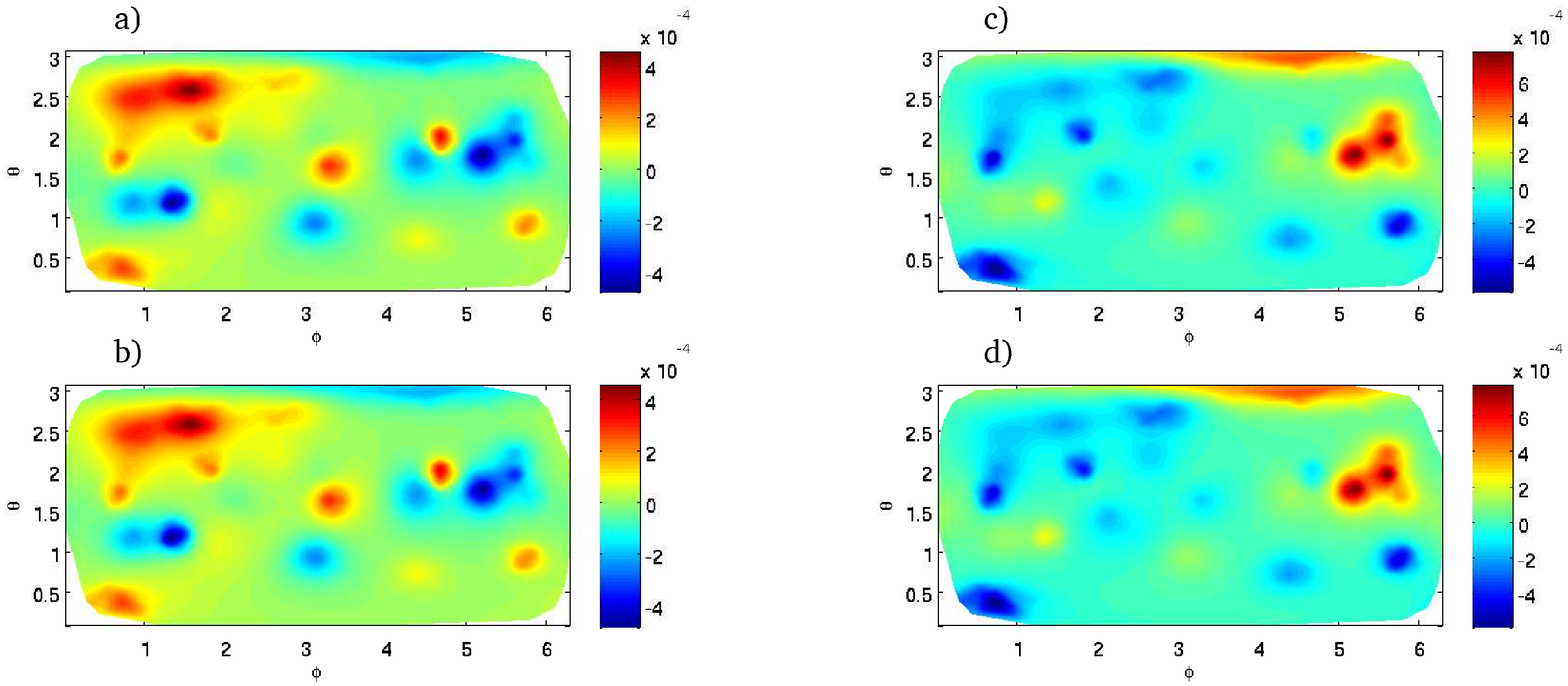}
}
\caption
{\label{fig7}
Polarization charge density (color-coded) (in units of $e/\sigma^{2}$) as a function of $\theta$ and $\phi$ 
induced on the dielectric sphere from a set of point charges around it (see Fig.~\ref{fig6}).
The permittivity inside the sphere is $\epsilon_{\textrm{in}}$ and the permittivity outside the sphere equals $\epsilon_{\textrm{out}}$. 
The left column represents the case when $\epsilon_{\textrm{in}} = 35$ and $\epsilon_{\textrm{out}} = 80$, while the right column shows the inverse case of
$\epsilon_{\textrm{in}} = 80$ and $\epsilon_{\textrm{out}} = 35$.
a) and c) Results from the numerical minimization of the functional $\mathscr{F}[\omega_{\mathbf{s}}]$.
b) and d) Exact results.
}
\end{figure*}
In Fig.~\ref{fig5} we show the polarization charge density for the single test charge problem. 
We place the positive unit charge on the $z$-axis at a distance of $d=12\sigma$ from the center of the sphere. 
Due to the inherent symmetry associated with this problem, the induced density is only a function of the $\theta$ variable. 
In addition, the density profile is symmetric around the $\theta = \pi$ point and hence the results are shown for 
$\theta \in [0,\pi]$. We compute the density
for the case when $\epsilon_{\textrm{in}}=35$, $\epsilon_{\textrm{out}}=80$ and also for the inverse problem where
$\epsilon_{\textrm{in}}=80$, $\epsilon_{\textrm{out}}=35$. 
As is evident from Fig.~\ref{fig5}, our numerical results
(red triangles) agree very well with the exact results (green circles) for both the cases studied.
We observe that in the first case ($\epsilon_{\textrm{in}} < \epsilon_{\textrm{out}}$), the density induced 
on the portion of the interface that is nearest to the charge (low $\theta$ values) is positive, 
while in the latter case ($\epsilon_{\textrm{in}} > \epsilon_{\textrm{out}}$) it is negative. 
Also, in both cases, the sign of the induced density flips at some value of $\theta$ (see inset in Fig.~\ref{fig5}). 
Furthermore, in either case, the magnitude of the induced density falls rapidly in the beginning 
as the angle $\theta$ increases. 
All these observations, which  are consistent with basic electrostatics principles, suggest that the test charge 
will repel away from the dielectric sphere when the dielctric constant of the latter is lower than the medium in which
the test charge is embedded. Otherwise, the test charge will be attracted towards the interface. 

We now present our findings for the case of many charges near the spherical interface. 
We consider 10 positive and 10 negative monovalent ions inside the sphere and the same outside. 
Thus, the total number of ions equals 40.
The positions of the ions are chosen at random and the ions remain fixed at their locations. 
As the system is electroneutral in each dielectric, the net induced 
charge on the sphere is 0.
In Fig.~\ref{fig7}(a) we show our simulation results for the polarization charge density 
at the interface as a function of the angles $\theta$ and $\phi$ for the case: 
$\epsilon_{\textrm{in}}=35$, $\epsilon_{\textrm{out}}=80$.
We refer to this graph as the polarization map.
For this case, the point charge present inside (outside) the dielectric induces
a charge of opposite (same) sign on the interface boundary closest to it.
The regions of intense red (highly positive) or intense blue (highly negative) on the polarization map suggest the 
presence of an ion or many ions near the interface at the corresponding $\theta,\phi$ location. 
Fig.~\ref{fig7}(b) shows the exact values of the induced density for this system. 
It is clear that the results from the numerical minimization of our functional are in excellent 
agreement with the exact results. 

In Fig.~\ref{fig7}(c) we show the polarization map for the same system as above, 
but with the dielectric media switched. 
Thus, for this case $\epsilon_{\textrm{in}}=80$ and $\epsilon_{\textrm{out}}=35$. 
We observe that by and large this map looks like the ``image'' of Fig.~\ref{fig7}(a).
Regions with more positive (red) induced charge
in Fig.~\ref{fig7}(c) are the ones that were highly negative (blue) in Fig.~\ref{fig7}(a)
and vice versa. We indeed expect this as now, in direct contrast to before,
an ion inside (outside) the dielectric induces a charge of the same (opposite) 
sign on the interface boundary closest to it.
Once again, comparison with exact results in Fig.~\ref{fig7}(d) 
confirms the accuracy of the numerical minimization procedure.

We note that since our functional is an energy functional, the above described numerical 
minimization procedure for the case when the ions are static, can be suitably modified 
to incorporate the scenario when ions are moving, like in a conventional MD simulation. 
Results from such a dynamical optimization procedure were presented in Ref.~\onlinecite{shortpaper}, 
along with a brief description of the method itself. 

\section{Conclusion}\label{sec:conclusion}
We have presented a variational formulation of electrostatics specifically designed to treat the problem of 
dielectric heterogeneities in charged systems.
Assuming only the condition of linear response, we constructed an energy functional 
that employs the polarization charge density as its sole variational field.
This functional is applicable for any configuration
of free charges and arbitrary spatial dependence of the dielectric response. 
We discussed in some depth the basic structure of our functional, drawing comparisons with past functionals and
revealing how more energy functionals can be constructed using our variational approach. 

Next, we focused on the important case of uniform dielectrics separated by sharp interfaces.
We showed that under this piecewise-uniform dielectric response, our functional reduces to a functional
of only the surface polarization charge density.  Such a reduction of the 3-dimensional electrostatic problem to a 2-dimensional one 
has many advantages from a computational perspective.
We then obtained the specific expressions for this reduced functional, and subsequently the 
induced charge density,
for the case of a point charge near a planar interface and for a point charge near a spherical dielectric.
Finally, in the view of applying our approach to more complicated systems, we discussed the numerical implementation 
of our minimizing variational principle for a system exhibiting piecewise-uniform dielectric response.
We illustrated this procedure for a system of monovalent ions near a spherical dielectric; obtaining
the polarization charge density induced on the interface and finding excellent 
agreement with exact results.

Due to the fact that $\mathscr{F}[\omega]$ is an energy functional, its minimization can be carried out 
in conjunction with the update 
of the ionic configuration. In Ref.~\onlinecite{shortpaper} we demonstrated such a dynamical minimization method.
This is of tremendous significance with regards to MD simulations of ions in heterogeneous media, as the explicit solution 
of the Poisson equation at each step is avoided. 
Detailed explorations of the dynamical optimization of our functional and the associated MD simulations
investigating diverse systems such as charged colloidal dispersions and liquid-liquid emulsions
will be the subject of a future study. 

\begin{acknowledgments}
V.J. thanks R. Sknepnek for many useful discussions.
V.J. was funded by the Department of Defense Research and Engineering (DDR\&E) and the Air Force Office of Scientific Research (AFOSR) under Award No. FA9550-10-1-0167 
and F.J.S. was funded by the NSF grant numbers DMR-0805330 and DMR-0907781.
\end{acknowledgments}

\appendix
\section{Extremal behavior of $\mathscr{F}[\omega]$}\label{sec:ext}
In this Appendix we investigate the extremal properties of $\mathscr{F}[\omega]$. 
First, we derive the condition for which $\mathscr{F}[\omega]$ is an extremum.
Next, we prove that at its extremum the functional gives the true electrostatic energy. 
And finally, we show that the functional becomes a minimum at its extremum.
\subsection{Extremum condition for $\mathscr{F}[\omega]$}\label{sec:ext-cond}
The derivation of the extremum condition for $\mathscr{F}[\omega]$ begins by  
recording how much the functional changes when the function $\omega$ is changed by an arbitrary small amount $\delta\omega$.
We use Eq.~\eqref{eq:fnal} to compute $\mathscr{F}[\omega+\delta\omega]$, retaining terms up to first order. The original functional $\mathscr{F}[\omega]$
is then subtracted from the result giving the first variation $\delta \mathscr{F} = \mathscr{F}[\omega+\delta\omega]-\mathscr{F}[\omega]$. 
Employing standard vector calculus identities \cite{arfken} wherever necessary and using Dirichlet boundary condition to make the surface integrals vanish 
by invoking the boundary at infinity,
we find the first variation $\delta\mathscr{F}$ to be
\begin{equation}\begin{split}\label{eq:varf}
\delta \mathscr{F} =& \int\delta\omega_{\mathbf{r}}\int G_{\mathbf{r},\mathbf{r'}}\times \\
&\nabla\cdot\left(\chi_{\mathbf{r'}}\nabla\int G_{\mathbf{r'},\mathbf{r''}}
\left(\Omega_{\mathbf{r''}} - \omega_{\mathbf{r''}}\right) d^{3}r''\right) d^{3}r'd^{3}r,
\end{split}
\end{equation}
where we have suppressed the functional part of the notation for $\Omega$ for brevity.
By definition, at the point of extremum, the first variation $\delta \mathscr{F}$ must vanish 
for an arbitrary $\delta\omega$. 
We see from \eqref{eq:varf} that this is only true if the following condition holds:
\begin{equation}\label{eq:extcond1}
\int G_{\mathbf{r},\mathbf{r'}}\nabla\cdot\left(\chi_{\mathbf{r'}}\nabla\int G_{\mathbf{r'},\mathbf{r''}}
\left(\Omega_{\mathbf{r''}} - \omega_{\mathbf{r''}}\right) d^{3}r''\right) d^{3}r' = 0.
\end{equation}

We now simplify Eq.~\eqref{eq:extcond1}.
Operating on both sides of \eqref{eq:extcond1} with the Laplacian operator and using \eqref{eq:deltafn} we obtain
\begin{equation}\label{eq:extcond2}
\nabla\cdot\left( \chi_{\mathbf{r}}\nabla\int G_{\mathbf{r},\mathbf{r'}}\left( \Omega_{\mathbf{r'}} -\omega_{\mathbf{r'}} \right)d^{3}r' \right)= 0,
\end{equation}
where towards the end we replaced the dummy variable $\mathbf{r''}$ with $\mathbf{r'}$.
It is useful to introduce  
\begin{equation}\label{eq:f}
f(\mathbf{r})= \int G_{\mathbf{r},\mathbf{r'}}\left( \Omega_{\mathbf{r'}} -\omega_{\mathbf{r'}} \right)d^{3}r',
\end{equation}
using which, Eq.~\eqref{eq:extcond2} can be written as 
\begin{equation}\label{eq:extcond3}
\nabla\cdot\left( \chi_{\mathbf{r}}\nabla f_{\mathbf{r}} \right)= 0.
\end{equation}
Multiplying both sides of \eqref{eq:extcond3} by $f(\mathbf{r})$ and integrating over whole space we obtain:
\begin{equation}\label{eq:extcond4}
\int f_{\mathbf{r}}\nabla\cdot\left( \chi_{\mathbf{r}}\nabla f_{\mathbf{r}} \right)d^{3}r= 0.
\end{equation}
Integrating by parts and employing DBC, we transform the above integral into 
\begin{equation}\label{extcond5}
\int\chi_{\mathbf{r}}\left|\nabla f_{\mathbf{r}}\right|^{2}d^{3}r= 0.
\end{equation}
It is clear that since $\chi(\mathbf{r})$ is always non-negative the integrand in the above equation is always non-negative. This means that the only way
the integral is zero is if the integrand is identically zero at all points, which implies $\nabla f(\mathbf{r}) = 0$ or $f(\mathbf{r})$ is a constant.
(Strictly speaking, the integrand can be zero without requiring that 
the gradient of $f$ vanishes: this happens when $\chi(\mathbf{r})$ vanishes at \emph{all} points.
But this situation represents the presence of free space everywhere, and in that case our functional becomes independent of $\omega$.)
Using \eqref{eq:f} to expand $f$, we thus obtain the equality:
\begin{equation}\label{eq:extcond5}
\int G_{\mathbf{r},\mathbf{r'}}\left( \Omega_{\mathbf{r'}} -\omega_{\mathbf{r'}} \right)d^{3}r'= c,
\end{equation}
where $c$ is some constant. Operating with the Laplacian on both sides of \eqref{eq:extcond5} and re-employing \eqref{eq:deltafn} we get
\begin{equation}\label{eq:shortextcond}
\omega_{\mathbf{r}} -\Omega_{\mathbf{r}}= 0,
\end{equation}
which, after expanding out the function $\Omega$ using \eqref{eq:Omega}, becomes
\begin{equation}\label{eq:extcond}
\omega_{\mathbf{r}}= 
\nabla\cdot\left( \chi_{\mathbf{r}}\nabla\int G_{\mathbf{r},\mathbf{r'}}\left(\rho_{\mathbf{r'}} + \omega_{\mathbf{r'}} \right) d^{3}r'\right).
\end{equation}
Equation \eqref{eq:extcond} gives the extremum condition for $\mathscr{F}[\omega]$. 
It is clear from the definition of $\omega$, the condition of linear response, and basic 
laws of electrostatics
that the right iterative relation for $\omega$ is obtained from the process of extremizing $\mathscr{F}[\omega]$.

\subsection{Value of $\mathscr{F}[\omega]$ at the extremum}
As a first step towards proving that the functional $\mathscr{F}[\omega]$ is an energy functional, we investigate here its value at extremum. 
Let $\bar{\omega}$ be the function that extremizes the functional $\mathscr{F}[\omega]$. The results of the last section show that $\bar{w}$ must satisfy
\eqref{eq:shortextcond} and so we obtain: 
\begin{equation}\label{eq:barw}
\bar{\omega}_{\mathbf{r}} - \Omega_{\mathbf{r}}[\bar{\omega}] =  0.
\end{equation}
To evaluate the value of $\mathscr{F}[\omega]$ at the extremum point 
we let $\omega = \bar{\omega}$ in \eqref{eq:fnal}, thus obtaining
\begin{equation}\begin{split}\label{eq:extfnal}
\mathscr{F}[\bar{\omega}]&=\frac{1}{2}\iint \rho_{\mathbf{r}}G_{\mathbf{r},\mathbf{r'}}
\left(\rho_{\mathbf{r'}}+\Omega_{\mathbf{r'}}[\bar{\omega}]\right) d^{3}r'd^{3}r\\
&- \frac{1}{2}\iint \Omega_{\mathbf{r}}[\bar{\omega}] G_{\mathbf{r},\mathbf{r'}}
\left(\bar{\omega}_{\mathbf{r'}} - \Omega_{\mathbf{r'}}[\bar{\omega}] \right) d^{3}r'd^{3}r.
\end{split}\end{equation}
Using \eqref{eq:barw}, the second double integral in the above equation vanishes and we obtain
\begin{equation}\begin{split}\label{eq:extfnal1}
\mathscr{F}[\bar{\omega}]&=\frac{1}{2}\iint \rho_{\mathbf{r}}G_{\mathbf{r},\mathbf{r'}}
\left(\rho_{\mathbf{r'}} + \bar{\omega}\right) d^{3}r'd^{3}r.
\end{split}\end{equation}

As we noted earlier, the function $\psi$ given by \eqref{eq:varomega} coincides with the electrostatic potential at the point of extremum. Thus, using \eqref{eq:varomega} we obtain
the following expression for the true electrostatic potential:
\begin{equation}\label{eq:pot}
\phi_{\mathbf{r}} = \int G_{\mathbf{r},\mathbf{r'}} \left(\rho_{\mathbf{r'}} + \bar{\omega}_{\mathbf{r'}}\right) d^{3}r'.
\end{equation}
Using \eqref{eq:pot}, the extremum value given in Eq.~\eqref{eq:extfnal1} becomes
\begin{equation}\begin{split}\label{eq:extfnal2}
\mathscr{F}[\bar{\omega}]&=\frac{1}{2}\int \rho_{\mathbf{r}}\phi_{\mathbf{r}}d^{3}r.
\end{split}\end{equation}

The expression on the right hand side of the above equation is the standard expression for the electrostatic energy, equivalent to 
$\frac{1}{8\pi}\int \epsilon\left(\mathbf{r}\right)\left|\mathbf{E}\left(\mathbf{r}\right)\right|^{2} d^{3}r$. 
Hence, the extremum value of $\mathscr{F}[\omega]$ gives the true electrostatic energy of the system.

\subsection{Proof that the extremum is a minimum}
To complete the proof that $\mathscr{F}[\omega]$ is an energy functional we now show that $\mathscr{F}[\omega]$ becomes a minimum at its extremum.
This begins by analyzing the terms in the variation of $\mathscr{F}[\omega]$ that are of second order in $\delta\omega$, terms which we ignored during the derivation of the extremum condition. 
If this second order change is shown to be positive then we would have proven that our functional becomes a minimum at the extremum point. 
It is clear from \eqref{eq:fnal} that the terms in $\delta\mathscr{F} = \mathscr{F}[\omega+\delta\omega] - \mathscr{F}[\omega]$ 
that are quadratic in $\delta\omega$ come only from the second double integral in \eqref{eq:fnal}, and we obtain
\begin{equation}\begin{split}\label{eq:proofmin2}
\delta^{2} \mathscr{F}=\frac{1}{2}\iint \delta\Omega_{\mathbf{r}} &G_{\mathbf{r},\mathbf{r'}} \delta\Omega_{\mathbf{r'}} d^{3}r'd^{3}r\\
&- \frac{1}{2}\iint \delta\Omega_{\mathbf{r}} G_{\mathbf{r},\mathbf{r'}} \delta\omega_{\mathbf{r'}} d^{3}r'd^{3}r,
\end{split}\end{equation}
where $\delta\Omega$ is given by 
\begin{equation}\label{eq:deltaO}
 \delta\Omega = \nabla\cdot\left( \chi_{\mathbf{r}}\nabla\int G_{\mathbf{r},\mathbf{r'}}\delta\omega_{\mathbf{r'}} d^{3}r'\right),
\end{equation}
and $\delta^{2}\mathscr{F}$ denotes the second order variation in $\mathscr{F}[\omega]$. 
We focus on the first double integral in \eqref{eq:proofmin2}. 
Integrating by parts and using the basic property of Green's function, namely, Eq.~\eqref{eq:deltafn}, the following identity can be derived:
\begin{equation}\begin{split}\label{eq:identity}
4\pi\iint h_{\mathbf{r}} G_{\mathbf{r},\mathbf{r'}} h_{\mathbf{r'}} d^{3}r'd^{3}r = \int\left|\nabla\int G_{\mathbf{r},\mathbf{r'}} h_{\mathbf{r'}} d^{3}r'\right|^{2} d^{3}r,
\end{split}\end{equation}
where $h$ is an arbitrary function. In deriving the above relation we invoked DBC for similar purposes as we have done before.
Using this identity with $h=\delta\Omega$, the first term in \eqref{eq:proofmin2} transforms to
\begin{equation}\begin{split}\label{eq:proofmin3}
\frac{1}{2}\iint \delta\Omega_{\mathbf{r}} &G_{\mathbf{r},\mathbf{r'}} \delta\Omega_{\mathbf{r'}} d^{3}r'd^{3}r\\
&=\frac{1}{8\pi}\int \left| \nabla\int G_{\mathbf{r},\mathbf{r'}}\delta\Omega_{\mathbf{r'}} d^{3}r' \right|^{2} d^{3}r.
\end{split}\end{equation}
We note that the right hand side of the above equation is always positive. 

We next probe the second term in \eqref{eq:proofmin2}.
Expanding $\delta\Omega$ using \eqref{eq:deltaO}, the second term becomes
\begin{equation}\begin{split}\label{eq:proofmin4}
&\frac{1}{2}\iint \delta\Omega_{\mathbf{r}} G_{\mathbf{r},\mathbf{r'}} \delta\omega_{\mathbf{r'}} d^{3}r'd^{3}r=\\
&\frac{1}{2}\int \nabla\cdot\left( \chi_{\mathbf{r}}\nabla \int G_{\mathbf{r},\mathbf{r''}}\delta\omega_{\mathbf{r''}}d^{3}r'' \right)
\int G_{\mathbf{r},\mathbf{r'}}\delta \omega_{\mathbf{r'}} d^{3}r'd^{3}r.
\end{split}\end{equation}
Integrating by parts and employing DBC we transform the right hand side of \eqref{eq:proofmin4} into 
a dot product of two gradients, as in
\begin{equation}\begin{split}\label{eq:proofmin6}
&\frac{1}{2}\iint \delta\Omega_{\mathbf{r}} G_{\mathbf{r},\mathbf{r'}} \delta\omega_{\mathbf{r'}} d^{3}r'd^{3}r=\\
&-\frac{1}{2}\int \chi_{\mathbf{r}}\nabla\int G_{\mathbf{r},\mathbf{r''}}\delta\omega_{\mathbf{r''}}d^{3}r''\cdot
\nabla\int G_{\mathbf{r},\mathbf{r'}}\delta \omega_{\mathbf{r'}} d^{3}r'd^{3}r,
\end{split}\end{equation}
which is equivalent to 
\begin{equation}\begin{split}\label{eq:proofmin8}
\frac{1}{2}\iint \delta\Omega_{\mathbf{r}} &G_{\mathbf{r},\mathbf{r'}} \delta\omega_{\mathbf{r'}} d^{3}r'd^{3}r=\\
&-\frac{1}{2}\int \chi_{\mathbf{r}}\left|\nabla\int G_{\mathbf{r},\mathbf{r'}}\delta\omega_{\mathbf{r'}}d^{3}r'\right|^{2}d^{3}r
\end{split}\end{equation}

The two double integrals in \eqref{eq:proofmin2} can now be replaced with expressions obtained in Eqs. \eqref{eq:proofmin3} and \eqref{eq:proofmin8}.
Doing so gives the following for the second variation:
\begin{equation}\begin{split}\label{eq:secondvar}
\delta^{2} \mathscr{F}=\frac{1}{8\pi}&\int \left| \nabla\int G_{\mathbf{r},\mathbf{r'}}\delta\Omega_{\mathbf{r'}} d^{3}r' \right|^{2} d^{3}r\\
&+ \frac{1}{2}\int \chi_{\mathbf{r}}\left|\nabla\int G_{\mathbf{r},\mathbf{r'}}\delta\omega_{\mathbf{r'}}d^{3}r'\right|^{2}d^{3}r.
\end{split}\end{equation}
Since $\chi(\mathbf{r})$ is non-negative everywhere, it is clear that both the terms on the right hand side of Eq.~\eqref{eq:secondvar} are always positive,
implying $\delta^{2}\mathscr{F} > 0$, thus completing the proof. 

\section{Point charges in uniform dielectric}\label{sec:app}
In this short appendix we apply our variational principle to the simplest case of a uniform dielectric. 
We derive the expression of our functional for this particular case and also obtain the induced density as 
a result of the minimization of the functional.

For a uniform dielectric $\chi(\mathbf{r}) = \chi_{u}$, where $\chi_{u}$ is a constant.
Employing this expression for $\chi(\mathbf{r})$ in \eqref{eq:Omega} and using \eqref{eq:greensfn}, we obtain
 \begin{equation}\label{eq:uOmega}
\Omega_{\mathbf{r}}[\omega] = -4\pi\chi_{u}\left( \rho_{\mathbf{r}} + \omega_{\mathbf{r}} \right).
\end{equation}
Substituting $\Omega_{\mathbf{r}}[\omega]$ from \eqref{eq:uOmega} in \eqref{eq:fnal} transforms the latter equation into
\begin{equation}\begin{split}\label{eq:ufnal}
\mathscr{F}_{\textrm{\tiny{U}}}[\omega]=
\frac{1}{2}&\left(\epsilon_{u}^{2} - 3\epsilon_{u} + 3\right)\iint \rho_{\mathbf{r}}G_{\mathbf{r},\mathbf{r'}}
\rho_{\mathbf{r'}} d^{3}r'd^{3}r \\
+ &\left(\epsilon_{u} - 1\right)^{2}\iint 
\rho_{\mathbf{r}}G_{\mathbf{r},\mathbf{r'}}\omega_{\mathbf{r'}}d^{3}r'd^{3}r \\
+ &\frac{1}{2}\epsilon_{u}\left(\epsilon_{u} - 1\right)\iint \omega_{\mathbf{r}}
G_{\mathbf{r},\mathbf{r'}}\omega_{\mathbf{r'}} d^{3}r'd^{3}r,
\end{split}
\end{equation}
where we have expressed the resulting functional in terms of the uniform permittivity $\epsilon_{u}$, which is connected 
to $\chi_{u}$ via the relation $\epsilon_{u} = 1 + 4\pi\chi_{u}$. 
Equation \eqref{eq:ufnal} gives the expression of our functional for the case of point charges in the presence of uniform dielectric response.

Let us now derive the expression for the density of induced charges in this case. 
From elementary electrostatics we expect that the induced charges are only to be found at the location of the free charges.
The first variation of the functional in \eqref{eq:ufnal} is:
\begin{equation}\begin{split}\label{eq:ufv}
\delta\mathscr{F}_{\textrm{\tiny{U}}}
=&\left(\epsilon_{u} - 1\right)^{2}\iint \rho_{\mathbf{r}} 
G_{\mathbf{r},\mathbf{r'}} \delta\omega_{\mathbf{r'}}d^{3}r'd^{3}r\\
&+\epsilon_{u}\left(\epsilon_{u} - 1\right)
\iint\omega_{\mathbf{r}} G_{\mathbf{r},\mathbf{r'}} \delta\omega_{\mathbf{r'}} d^{3}r'd^{3}r.
\end{split}
\end{equation}
For $\delta\mathscr{F}_{\textrm{\tiny{U}}}$ to vanish for any $\delta\omega$,
it is clear from \eqref{eq:ufv} that the following must be true:
\begin{equation}\begin{split}\label{eq:uextcond}
\int G_{\mathbf{r},\mathbf{r'}}\left(\left(\epsilon_{u} - 1\right)^{2}\rho_{\mathbf{r'}}
+\epsilon_{u}\left(\epsilon_{u} - 1\right)\omega_{\mathbf{r'}}\right)d^{3}r' = 0,
\end{split}
\end{equation}
Applying the Laplacian on both sides of \eqref{eq:uextcond} and employing \eqref{eq:deltafn}, we obtain
\begin{equation}
\left(\epsilon_{u} - 1\right)^{2}\rho_{\mathbf{r}} + \epsilon_{u}\left(\epsilon_{u} - 1\right)\omega_{\mathbf{r}} = 0,
\end{equation}
which simplifies to
\begin{equation}\begin{split}\label{eq:ucw}
\omega_{\mathbf{r}} = -\frac{\epsilon_{u} - 1}{\epsilon_{u}}\rho_{\mathbf{r}}.
\end{split}
\end{equation}
Equation \eqref{eq:ucw} is indeed the standard expression for the induced charge density for the 
case of a uniform dielectric.


\begin{thebibliography}{45}%
\makeatletter
\providecommand \@ifxundefined [1]{%
 \@ifx{#1\undefined}
}%
\providecommand \@ifnum [1]{%
 \ifnum #1\expandafter \@firstoftwo
 \else \expandafter \@secondoftwo
 \fi
}%
\providecommand \@ifx [1]{%
 \ifx #1\expandafter \@firstoftwo
 \else \expandafter \@secondoftwo
 \fi
}%
\providecommand \natexlab [1]{#1}%
\providecommand \enquote  [1]{``#1''}%
\providecommand \bibnamefont  [1]{#1}%
\providecommand \bibfnamefont [1]{#1}%
\providecommand \citenamefont [1]{#1}%
\providecommand \href@noop [0]{\@secondoftwo}%
\providecommand \href [0]{\begingroup \@sanitize@url \@href}%
\providecommand \@href[1]{\@@startlink{#1}\@@href}%
\providecommand \@@href[1]{\endgroup#1\@@endlink}%
\providecommand \@sanitize@url [0]{\catcode `\\12\catcode `\$12\catcode
  `\&12\catcode `\#12\catcode `\^12\catcode `\_12\catcode `\%12\relax}%
\providecommand \@@startlink[1]{}%
\providecommand \@@endlink[0]{}%
\providecommand \url  [0]{\begingroup\@sanitize@url \@url }%
\providecommand \@url [1]{\endgroup\@href {#1}{\urlprefix }}%
\providecommand \urlprefix  [0]{URL }%
\providecommand \Eprint [0]{\href }%
\providecommand \doibase [0]{http://dx.doi.org/}%
\providecommand \selectlanguage [0]{\@gobble}%
\providecommand \bibinfo  [0]{\@secondoftwo}%
\providecommand \bibfield  [0]{\@secondoftwo}%
\providecommand \translation [1]{[#1]}%
\providecommand \BibitemOpen [0]{}%
\providecommand \bibitemStop [0]{}%
\providecommand \bibitemNoStop [0]{.\EOS\space}%
\providecommand \EOS [0]{\spacefactor3000\relax}%
\providecommand \BibitemShut  [1]{\csname bibitem#1\endcsname}%
\let\auto@bib@innerbib\@empty
%</preamble>
\bibitem [{\citenamefont {Honig}\ and\ \citenamefont {Nicholls}(1995)}]{honig}%
  \BibitemOpen
  \bibfield  {author} {\bibinfo {author} {\bibfnamefont {B.}~\bibnamefont
  {Honig}}\ and\ \bibinfo {author} {\bibfnamefont {A.}~\bibnamefont
  {Nicholls}},\ }\href@noop {} {\bibfield  {journal} {\bibinfo  {journal}
  {Science}\ }\textbf {\bibinfo {volume} {268}},\ \bibinfo {pages} {1144}
  (\bibinfo {year} {1995})}\BibitemShut {NoStop}%
\bibitem [{\citenamefont {Perutz}(1978)}]{perutz}%
  \BibitemOpen
  \bibfield  {author} {\bibinfo {author} {\bibfnamefont {M.}~\bibnamefont
  {Perutz}},\ }\href@noop {} {\bibfield  {journal} {\bibinfo  {journal}
  {Science}\ }\textbf {\bibinfo {volume} {201}},\ \bibinfo {pages} {1187}
  (\bibinfo {year} {1978})}\BibitemShut {NoStop}%
\bibitem [{\citenamefont {Clapham}(2007)}]{clapham}%
  \BibitemOpen
  \bibfield  {author} {\bibinfo {author} {\bibfnamefont {D.~E.}\ \bibnamefont
  {Clapham}},\ }\href@noop {} {\bibfield  {journal} {\bibinfo  {journal}
  {Cell}\ }\textbf {\bibinfo {volume} {131}},\ \bibinfo {pages} {1047 }
  (\bibinfo {year} {2007})}\BibitemShut {NoStop}%
\bibitem [{\citenamefont {Levin}(2005)}]{levin1}%
  \BibitemOpen
  \bibfield  {author} {\bibinfo {author} {\bibfnamefont {Y.}~\bibnamefont
  {Levin}},\ }\href@noop {} {\bibfield  {journal} {\bibinfo  {journal} {Physica
  A: Statistical Mechanics and its Applications}\ }\textbf {\bibinfo {volume}
  {352}},\ \bibinfo {pages} {43 } (\bibinfo {year} {2005})}\BibitemShut
  {NoStop}%
\bibitem [{\citenamefont {Cheng}\ \emph {et~al.}(2006)\citenamefont {Cheng},
  \citenamefont {Zhang}, \citenamefont {Libera}, \citenamefont {Olvera de~la
  Cruz},\ and\ \citenamefont {Bedzyk}}]{bedzyk}%
  \BibitemOpen
  \bibfield  {author} {\bibinfo {author} {\bibfnamefont {H.}~\bibnamefont
  {Cheng}}, \bibinfo {author} {\bibfnamefont {K.}~\bibnamefont {Zhang}},
  \bibinfo {author} {\bibfnamefont {J.~A.}\ \bibnamefont {Libera}}, \bibinfo
  {author} {\bibfnamefont {M.}~\bibnamefont {Olvera de~la Cruz}}, \ and\
  \bibinfo {author} {\bibfnamefont {M.~J.}\ \bibnamefont {Bedzyk}},\
  }\href@noop {} {\bibfield  {journal} {\bibinfo  {journal} {Biophys J}\
  }\textbf {\bibinfo {volume} {90}},\ \bibinfo {pages} {1164} (\bibinfo {year}
  {2006})}\BibitemShut {NoStop}%
\bibitem [{\citenamefont {Rouzina}\ and\ \citenamefont
  {Bloomfield}(1996)}]{rouzina}%
  \BibitemOpen
  \bibfield  {author} {\bibinfo {author} {\bibfnamefont {I.}~\bibnamefont
  {Rouzina}}\ and\ \bibinfo {author} {\bibfnamefont {V.~A.}\ \bibnamefont
  {Bloomfield}},\ }\href@noop {} {\bibfield  {journal} {\bibinfo  {journal}
  {The Journal of Physical Chemistry}\ }\textbf {\bibinfo {volume} {100}},\
  \bibinfo {pages} {9977} (\bibinfo {year} {1996})}\BibitemShut {NoStop}%
\bibitem [{\citenamefont {Raspaud}\ \emph {et~al.}(1998)\citenamefont
  {Raspaud}, \citenamefont {Olvera de~la Cruz}, \citenamefont {Sikorav},\ and\
  \citenamefont {Livolant}}]{raspaud}%
  \BibitemOpen
  \bibfield  {author} {\bibinfo {author} {\bibfnamefont {E.}~\bibnamefont
  {Raspaud}}, \bibinfo {author} {\bibfnamefont {M.}~\bibnamefont {Olvera de~la
  Cruz}}, \bibinfo {author} {\bibfnamefont {J.}~\bibnamefont {Sikorav}}, \ and\
  \bibinfo {author} {\bibfnamefont {F.}~\bibnamefont {Livolant}},\ }\href@noop
  {} {\bibfield  {journal} {\bibinfo  {journal} {Biophys J}\ }\textbf {\bibinfo
  {volume} {74}},\ \bibinfo {pages} {381} (\bibinfo {year} {1998})}\BibitemShut
  {NoStop}%
\bibitem [{\citenamefont {van~der Heyden}\ \emph {et~al.}(2006)\citenamefont
  {van~der Heyden}, \citenamefont {Stein}, \citenamefont {Besteman},
  \citenamefont {Lemay},\ and\ \citenamefont {Dekker}}]{charge_inversion}%
  \BibitemOpen
  \bibfield  {author} {\bibinfo {author} {\bibfnamefont {F.~H.~J.}\
  \bibnamefont {van~der Heyden}}, \bibinfo {author} {\bibfnamefont
  {D.}~\bibnamefont {Stein}}, \bibinfo {author} {\bibfnamefont
  {K.}~\bibnamefont {Besteman}}, \bibinfo {author} {\bibfnamefont {S.~G.}\
  \bibnamefont {Lemay}}, \ and\ \bibinfo {author} {\bibfnamefont
  {C.}~\bibnamefont {Dekker}},\ }\href@noop {} {\bibfield  {journal} {\bibinfo
  {journal} {Phys. Rev. Lett.}\ }\textbf {\bibinfo {volume} {96}},\ \bibinfo
  {pages} {224502} (\bibinfo {year} {2006})}\BibitemShut {NoStop}%
\bibitem [{\citenamefont {Wernersson}\ \emph {et~al.}(2010)\citenamefont
  {Wernersson}, \citenamefont {Kjellander},\ and\ \citenamefont
  {Lyklema}}]{charge_inversion2}%
  \BibitemOpen
  \bibfield  {author} {\bibinfo {author} {\bibfnamefont {E.}~\bibnamefont
  {Wernersson}}, \bibinfo {author} {\bibfnamefont {R.}~\bibnamefont
  {Kjellander}}, \ and\ \bibinfo {author} {\bibfnamefont {J.}~\bibnamefont
  {Lyklema}},\ }\href@noop {} {\bibfield  {journal} {\bibinfo  {journal} {The
  Journal of Physical Chemistry C}\ }\textbf {\bibinfo {volume} {114}},\
  \bibinfo {pages} {1849} (\bibinfo {year} {2010})}\BibitemShut {NoStop}%
\bibitem [{\citenamefont {Solis}\ \emph {et~al.}(2011)\citenamefont {Solis},
  \citenamefont {Vernizzi},\ and\ \citenamefont {Olvera de~la Cruz}}]{paco}%
  \BibitemOpen
  \bibfield  {author} {\bibinfo {author} {\bibfnamefont {F.~J.}\ \bibnamefont
  {Solis}}, \bibinfo {author} {\bibfnamefont {G.}~\bibnamefont {Vernizzi}}, \
  and\ \bibinfo {author} {\bibfnamefont {M.}~\bibnamefont {Olvera de~la
  Cruz}},\ }\href@noop {} {\bibfield  {journal} {\bibinfo  {journal} {Soft
  Matter}\ }\textbf {\bibinfo {volume} {7}},\ \bibinfo {pages} {1456} (\bibinfo
  {year} {2011})}\BibitemShut {NoStop}%
\bibitem [{\citenamefont {Bier}\ \emph {et~al.}(2008)\citenamefont {Bier},
  \citenamefont {Zwanikken},\ and\ \citenamefont {van Roij}}]{bier}%
  \BibitemOpen
  \bibfield  {author} {\bibinfo {author} {\bibfnamefont {M.}~\bibnamefont
  {Bier}}, \bibinfo {author} {\bibfnamefont {J.}~\bibnamefont {Zwanikken}}, \
  and\ \bibinfo {author} {\bibfnamefont {R.}~\bibnamefont {van Roij}},\
  }\href@noop {} {\bibfield  {journal} {\bibinfo  {journal} {Phys. Rev. Lett.}\
  }\textbf {\bibinfo {volume} {101}},\ \bibinfo {pages} {046104} (\bibinfo
  {year} {2008})}\BibitemShut {NoStop}%
\bibitem [{\citenamefont {Kung}\ \emph {et~al.}(2009)\citenamefont {Kung},
  \citenamefont {Solis},\ and\ \citenamefont {Olvera de~la Cruz}}]{kung}%
  \BibitemOpen
  \bibfield  {author} {\bibinfo {author} {\bibfnamefont {W.}~\bibnamefont
  {Kung}}, \bibinfo {author} {\bibfnamefont {F.~J.}\ \bibnamefont {Solis}}, \
  and\ \bibinfo {author} {\bibfnamefont {M.}~\bibnamefont {Olvera de~la
  Cruz}},\ }\href@noop {} {\bibfield  {journal} {\bibinfo  {journal} {The
  Journal of Chemical Physics}\ }\textbf {\bibinfo {volume} {130}},\ \bibinfo
  {eid} {044502} (\bibinfo {year} {2009})}\BibitemShut {NoStop}%
\bibitem [{\citenamefont {Wang}(2008)}]{wang}%
  \BibitemOpen
  \bibfield  {author} {\bibinfo {author} {\bibfnamefont {Z.-G.}\ \bibnamefont
  {Wang}},\ }\href@noop {} {\bibfield  {journal} {\bibinfo  {journal} {Journal
  of Theoretical and Computational Chemistry}\ }\textbf {\bibinfo {volume}
  {07}},\ \bibinfo {pages} {397} (\bibinfo {year} {2008})}\BibitemShut
  {NoStop}%
\bibitem [{\citenamefont {Vernizzi}\ and\ \citenamefont {Olvera de~la
  Cruz}(2007)}]{vernizzi}%
  \BibitemOpen
  \bibfield  {author} {\bibinfo {author} {\bibfnamefont {G.}~\bibnamefont
  {Vernizzi}}\ and\ \bibinfo {author} {\bibfnamefont {M.}~\bibnamefont {Olvera
  de~la Cruz}},\ }\href@noop {} {\bibfield  {journal} {\bibinfo  {journal}
  {Proceedings of the National Academy of Sciences}\ }\textbf {\bibinfo
  {volume} {104}},\ \bibinfo {pages} {18382} (\bibinfo {year}
  {2007})}\BibitemShut {NoStop}%
\bibitem [{\citenamefont {Mahalik}\ and\ \citenamefont
  {Muthukumar}(2012)}]{muthu}%
  \BibitemOpen
  \bibfield  {author} {\bibinfo {author} {\bibfnamefont {J.~P.}\ \bibnamefont
  {Mahalik}}\ and\ \bibinfo {author} {\bibfnamefont {M.}~\bibnamefont
  {Muthukumar}},\ }\href@noop {} {\bibfield  {journal} {\bibinfo  {journal}
  {The Journal of Chemical Physics}\ }\textbf {\bibinfo {volume} {136}},\
  \bibinfo {eid} {135101} (\bibinfo {year} {2012})}\BibitemShut {NoStop}%
\bibitem [{\citenamefont {Fischer}\ \emph {et~al.}(2008)\citenamefont
  {Fischer}, \citenamefont {Naji},\ and\ \citenamefont {Netz}}]{netz}%
  \BibitemOpen
  \bibfield  {author} {\bibinfo {author} {\bibfnamefont {S.}~\bibnamefont
  {Fischer}}, \bibinfo {author} {\bibfnamefont {A.}~\bibnamefont {Naji}}, \
  and\ \bibinfo {author} {\bibfnamefont {R.~R.}\ \bibnamefont {Netz}},\
  }\href@noop {} {\bibfield  {journal} {\bibinfo  {journal} {Phys. Rev. Lett.}\
  }\textbf {\bibinfo {volume} {101}},\ \bibinfo {pages} {176103} (\bibinfo
  {year} {2008})}\BibitemShut {NoStop}%
\bibitem [{\citenamefont {Jendrejack}\ \emph {et~al.}(2002)\citenamefont
  {Jendrejack}, \citenamefont {de~Pablo},\ and\ \citenamefont
  {Graham}}]{depablo}%
  \BibitemOpen
  \bibfield  {author} {\bibinfo {author} {\bibfnamefont {R.~M.}\ \bibnamefont
  {Jendrejack}}, \bibinfo {author} {\bibfnamefont {J.~J.}\ \bibnamefont
  {de~Pablo}}, \ and\ \bibinfo {author} {\bibfnamefont {M.~D.}\ \bibnamefont
  {Graham}},\ }\href@noop {} {\bibfield  {journal} {\bibinfo  {journal} {The
  Journal of Chemical Physics}\ }\textbf {\bibinfo {volume} {116}},\ \bibinfo
  {pages} {7752} (\bibinfo {year} {2002})}\BibitemShut {NoStop}%
\bibitem [{\citenamefont {Grass}\ and\ \citenamefont {Holm}(2009)}]{holm}%
  \BibitemOpen
  \bibfield  {author} {\bibinfo {author} {\bibfnamefont {K.}~\bibnamefont
  {Grass}}\ and\ \bibinfo {author} {\bibfnamefont {C.}~\bibnamefont {Holm}},\
  }\href@noop {} {\bibfield  {journal} {\bibinfo  {journal} {Soft Matter}\
  }\textbf {\bibinfo {volume} {5}},\ \bibinfo {pages} {2079} (\bibinfo {year}
  {2009})}\BibitemShut {NoStop}%
\bibitem [{\citenamefont {Guerrero-Garcia}\ \emph {et~al.}(2011)\citenamefont
  {Guerrero-Garcia}, \citenamefont {Gonzalez-Tovar},\ and\ \citenamefont
  {Olvera de~la Cruz}}]{ivan}%
  \BibitemOpen
  \bibfield  {author} {\bibinfo {author} {\bibfnamefont {G.~I.}\ \bibnamefont
  {Guerrero-Garcia}}, \bibinfo {author} {\bibfnamefont {E.}~\bibnamefont
  {Gonzalez-Tovar}}, \ and\ \bibinfo {author} {\bibfnamefont {M.}~\bibnamefont
  {Olvera de~la Cruz}},\ }\href@noop {} {\bibfield  {journal} {\bibinfo
  {journal} {The Journal of Chemical Physics}\ }\textbf {\bibinfo {volume}
  {135}},\ \bibinfo {eid} {054701} (\bibinfo {year} {2011})}\BibitemShut
  {NoStop}%
\bibitem [{\citenamefont {Hatlo}\ and\ \citenamefont {Lue}(2008)}]{lue1}%
  \BibitemOpen
  \bibfield  {author} {\bibinfo {author} {\bibfnamefont {M.~M.}\ \bibnamefont
  {Hatlo}}\ and\ \bibinfo {author} {\bibfnamefont {L.}~\bibnamefont {Lue}},\
  }\href@noop {} {\bibfield  {journal} {\bibinfo  {journal} {Soft Matter}\
  }\textbf {\bibinfo {volume} {4}},\ \bibinfo {pages} {1582} (\bibinfo {year}
  {2008})}\BibitemShut {NoStop}%
\bibitem [{\citenamefont {Sagui}\ and\ \citenamefont {Darden}(1999)}]{sagui1}%
  \BibitemOpen
  \bibfield  {author} {\bibinfo {author} {\bibfnamefont {C.}~\bibnamefont
  {Sagui}}\ and\ \bibinfo {author} {\bibfnamefont {T.}~\bibnamefont {Darden}},\
  }\href@noop {} {\bibfield  {journal} {\bibinfo  {journal} {Annu. Rev.
  Biophys. Biomol. Struct.}\ }\textbf {\bibinfo {volume} {28}},\ \bibinfo
  {pages} {155} (\bibinfo {year} {1999})}\BibitemShut {NoStop}%
\bibitem [{\citenamefont {Sagui}\ and\ \citenamefont {Darden}(2001)}]{sagui2}%
  \BibitemOpen
  \bibfield  {author} {\bibinfo {author} {\bibfnamefont {C.}~\bibnamefont
  {Sagui}}\ and\ \bibinfo {author} {\bibfnamefont {T.}~\bibnamefont {Darden}},\
  }\href@noop {} {\bibfield  {journal} {\bibinfo  {journal} {The Journal of
  Chemical Physics}\ }\textbf {\bibinfo {volume} {114}},\ \bibinfo {pages}
  {6578} (\bibinfo {year} {2001})}\BibitemShut {NoStop}%
\bibitem [{\citenamefont {Maggs}\ and\ \citenamefont
  {Rossetto}(2002)}]{maggs-rossetto}%
  \BibitemOpen
  \bibfield  {author} {\bibinfo {author} {\bibfnamefont {A.~C.}\ \bibnamefont
  {Maggs}}\ and\ \bibinfo {author} {\bibfnamefont {V.}~\bibnamefont
  {Rossetto}},\ }\href@noop {} {\bibfield  {journal} {\bibinfo  {journal}
  {Phys. Rev. Lett.}\ }\textbf {\bibinfo {volume} {88}},\ \bibinfo {pages}
  {196402} (\bibinfo {year} {2002})}\BibitemShut {NoStop}%
\bibitem [{\citenamefont {Rottler}\ and\ \citenamefont
  {Maggs}(2004)}]{rottler-maggs}%
  \BibitemOpen
  \bibfield  {author} {\bibinfo {author} {\bibfnamefont {J.}~\bibnamefont
  {Rottler}}\ and\ \bibinfo {author} {\bibfnamefont {A.~C.}\ \bibnamefont
  {Maggs}},\ }\href@noop {} {\bibfield  {journal} {\bibinfo  {journal} {Phys.
  Rev. Lett.}\ }\textbf {\bibinfo {volume} {93}},\ \bibinfo {pages} {170201}
  (\bibinfo {year} {2004})}\BibitemShut {NoStop}%
\bibitem [{\citenamefont {Sacanna}\ \emph {et~al.}(2007)\citenamefont
  {Sacanna}, \citenamefont {Kegel},\ and\ \citenamefont {Philipse}}]{sacanna}%
  \BibitemOpen
  \bibfield  {author} {\bibinfo {author} {\bibfnamefont {S.}~\bibnamefont
  {Sacanna}}, \bibinfo {author} {\bibfnamefont {W.~K.}\ \bibnamefont {Kegel}},
  \ and\ \bibinfo {author} {\bibfnamefont {A.~P.}\ \bibnamefont {Philipse}},\
  }\href@noop {} {\bibfield  {journal} {\bibinfo  {journal} {Phys. Rev. Lett.}\
  }\textbf {\bibinfo {volume} {98}},\ \bibinfo {pages} {158301} (\bibinfo
  {year} {2007})}\BibitemShut {NoStop}%
\bibitem [{\citenamefont {Marchi}\ \emph {et~al.}(2001)\citenamefont {Marchi},
  \citenamefont {Borgis}, \citenamefont {Levy},\ and\ \citenamefont
  {Ballone}}]{marchi}%
  \BibitemOpen
  \bibfield  {author} {\bibinfo {author} {\bibfnamefont {M.}~\bibnamefont
  {Marchi}}, \bibinfo {author} {\bibfnamefont {D.}~\bibnamefont {Borgis}},
  \bibinfo {author} {\bibfnamefont {N.}~\bibnamefont {Levy}}, \ and\ \bibinfo
  {author} {\bibfnamefont {P.}~\bibnamefont {Ballone}},\ }\href@noop {}
  {\bibfield  {journal} {\bibinfo  {journal} {The Journal of Chemical Physics}\
  }\textbf {\bibinfo {volume} {114}},\ \bibinfo {pages} {4377} (\bibinfo {year}
  {2001})}\BibitemShut {NoStop}%
\bibitem [{\citenamefont {Allen}\ \emph {et~al.}(2001)\citenamefont {Allen},
  \citenamefont {Hansen},\ and\ \citenamefont {Melchionna}}]{allen1}%
  \BibitemOpen
  \bibfield  {author} {\bibinfo {author} {\bibfnamefont {R.}~\bibnamefont
  {Allen}}, \bibinfo {author} {\bibfnamefont {J.-P.}\ \bibnamefont {Hansen}}, \
  and\ \bibinfo {author} {\bibfnamefont {S.}~\bibnamefont {Melchionna}},\
  }\href@noop {} {\bibfield  {journal} {\bibinfo  {journal} {Phys. Chem. Chem.
  Phys.}\ }\textbf {\bibinfo {volume} {3}},\ \bibinfo {pages} {4177} (\bibinfo
  {year} {2001})}\BibitemShut {NoStop}%
\bibitem [{\citenamefont {Messina}(2002)}]{messina1}%
  \BibitemOpen
  \bibfield  {author} {\bibinfo {author} {\bibfnamefont {R.}~\bibnamefont
  {Messina}},\ }\href@noop {} {\bibfield  {journal} {\bibinfo  {journal} {The
  Journal of Chemical Physics}\ }\textbf {\bibinfo {volume} {117}},\ \bibinfo
  {pages} {11062} (\bibinfo {year} {2002})}\BibitemShut {NoStop}%
\bibitem [{\citenamefont {Boda}\ \emph {et~al.}(2004)\citenamefont {Boda},
  \citenamefont {Gillespie}, \citenamefont {Nonner}, \citenamefont
  {Henderson},\ and\ \citenamefont {Eisenberg}}]{boda}%
  \BibitemOpen
  \bibfield  {author} {\bibinfo {author} {\bibfnamefont {D.}~\bibnamefont
  {Boda}}, \bibinfo {author} {\bibfnamefont {D.}~\bibnamefont {Gillespie}},
  \bibinfo {author} {\bibfnamefont {W.}~\bibnamefont {Nonner}}, \bibinfo
  {author} {\bibfnamefont {D.}~\bibnamefont {Henderson}}, \ and\ \bibinfo
  {author} {\bibfnamefont {B.}~\bibnamefont {Eisenberg}},\ }\href@noop {}
  {\bibfield  {journal} {\bibinfo  {journal} {Phys. Rev. E}\ }\textbf {\bibinfo
  {volume} {69}},\ \bibinfo {pages} {046702} (\bibinfo {year}
  {2004})}\BibitemShut {NoStop}%
\bibitem [{\citenamefont {Attard}(2003)}]{attard}%
  \BibitemOpen
  \bibfield  {author} {\bibinfo {author} {\bibfnamefont {P.}~\bibnamefont
  {Attard}},\ }\href@noop {} {\bibfield  {journal} {\bibinfo  {journal} {The
  Journal of Chemical Physics}\ }\textbf {\bibinfo {volume} {119}},\ \bibinfo
  {pages} {1365} (\bibinfo {year} {2003})}\BibitemShut {NoStop}%
\bibitem [{\citenamefont {Linse}(2008)}]{linse}%
  \BibitemOpen
  \bibfield  {author} {\bibinfo {author} {\bibfnamefont {P.}~\bibnamefont
  {Linse}},\ }\href@noop {} {\bibfield  {journal} {\bibinfo  {journal} {The
  Journal of Chemical Physics}\ }\textbf {\bibinfo {volume} {128}},\ \bibinfo
  {eid} {214505} (\bibinfo {year} {2008})}\BibitemShut {NoStop}%
\bibitem [{\citenamefont {Gan}\ and\ \citenamefont {Xu}(2011)}]{gan}%
  \BibitemOpen
  \bibfield  {author} {\bibinfo {author} {\bibfnamefont {Z.}~\bibnamefont
  {Gan}}\ and\ \bibinfo {author} {\bibfnamefont {Z.}~\bibnamefont {Xu}},\
  }\href@noop {} {\bibfield  {journal} {\bibinfo  {journal} {Phys. Rev. E}\
  }\textbf {\bibinfo {volume} {84}},\ \bibinfo {pages} {016705} (\bibinfo
  {year} {2011})}\BibitemShut {NoStop}%
\bibitem [{\citenamefont {Tyagi}\ \emph {et~al.}(2010)\citenamefont {Tyagi},
  \citenamefont {Suzen}, \citenamefont {Sega}, \citenamefont {Barbosa},
  \citenamefont {Kantorovich},\ and\ \citenamefont {Holm}}]{tyagi}%
  \BibitemOpen
  \bibfield  {author} {\bibinfo {author} {\bibfnamefont {S.}~\bibnamefont
  {Tyagi}}, \bibinfo {author} {\bibfnamefont {M.}~\bibnamefont {Suzen}},
  \bibinfo {author} {\bibfnamefont {M.}~\bibnamefont {Sega}}, \bibinfo {author}
  {\bibfnamefont {M.}~\bibnamefont {Barbosa}}, \bibinfo {author} {\bibfnamefont
  {S.~S.}\ \bibnamefont {Kantorovich}}, \ and\ \bibinfo {author} {\bibfnamefont
  {C.}~\bibnamefont {Holm}},\ }\href@noop {} {\bibfield  {journal} {\bibinfo
  {journal} {The Journal of Chemical Physics}\ }\textbf {\bibinfo {volume}
  {132}},\ \bibinfo {eid} {154112} (\bibinfo {year} {2010})}\BibitemShut
  {NoStop}%
\bibitem [{\citenamefont {dos Santos}\ \emph {et~al.}(2011)\citenamefont {dos
  Santos}, \citenamefont {Bakhshandeh},\ and\ \citenamefont {Levin}}]{santos}%
  \BibitemOpen
  \bibfield  {author} {\bibinfo {author} {\bibfnamefont {A.~P.}\ \bibnamefont
  {dos Santos}}, \bibinfo {author} {\bibfnamefont {A.}~\bibnamefont
  {Bakhshandeh}}, \ and\ \bibinfo {author} {\bibfnamefont {Y.}~\bibnamefont
  {Levin}},\ }\href@noop {} {\bibfield  {journal} {\bibinfo  {journal} {The
  Journal of Chemical Physics}\ }\textbf {\bibinfo {volume} {135}},\ \bibinfo
  {eid} {044124} (\bibinfo {year} {2011})}\BibitemShut {NoStop}%
\bibitem [{\citenamefont {Lipparini}\ \emph {et~al.}(2010)\citenamefont
  {Lipparini}, \citenamefont {Scalmani}, \citenamefont {Mennucci},
  \citenamefont {Cances}, \citenamefont {Caricato},\ and\ \citenamefont
  {Frisch}}]{lipparini}%
  \BibitemOpen
  \bibfield  {author} {\bibinfo {author} {\bibfnamefont {F.}~\bibnamefont
  {Lipparini}}, \bibinfo {author} {\bibfnamefont {G.}~\bibnamefont {Scalmani}},
  \bibinfo {author} {\bibfnamefont {B.}~\bibnamefont {Mennucci}}, \bibinfo
  {author} {\bibfnamefont {E.}~\bibnamefont {Cances}}, \bibinfo {author}
  {\bibfnamefont {M.}~\bibnamefont {Caricato}}, \ and\ \bibinfo {author}
  {\bibfnamefont {M.~J.}\ \bibnamefont {Frisch}},\ }\href@noop {} {\bibfield
  {journal} {\bibinfo  {journal} {The Journal of Chemical Physics}\ }\textbf
  {\bibinfo {volume} {133}},\ \bibinfo {eid} {014106} (\bibinfo {year}
  {2010})}\BibitemShut {NoStop}%
\bibitem [{\citenamefont {Jackson}(1999)}]{jackson}%
  \BibitemOpen
  \bibfield  {author} {\bibinfo {author} {\bibfnamefont {J.~D.}\ \bibnamefont
  {Jackson}},\ }\href@noop {} {\emph {\bibinfo {title} {Classical
  Electrodynamics}}},\ \bibinfo {edition} {3rd}\ ed.\ (\bibinfo  {publisher}
  {Wiley, New York},\ \bibinfo {year} {1999})\BibitemShut {NoStop}%
\bibitem [{\citenamefont {Schwinger}\ \emph {et~al.}(1998)\citenamefont
  {Schwinger}, \citenamefont {Deraad}, \citenamefont {Milton}, \citenamefont
  {Tsai},\ and\ \citenamefont {Norton}}]{schwinger}%
  \BibitemOpen
  \bibfield  {author} {\bibinfo {author} {\bibfnamefont {J.}~\bibnamefont
  {Schwinger}}, \bibinfo {author} {\bibfnamefont {L.}~\bibnamefont {Deraad}},
  \bibinfo {author} {\bibfnamefont {K.}~\bibnamefont {Milton}}, \bibinfo
  {author} {\bibfnamefont {W.}~\bibnamefont {Tsai}}, \ and\ \bibinfo {author}
  {\bibfnamefont {J.}~\bibnamefont {Norton}},\ }\href@noop {} {\emph {\bibinfo
  {title} {Classical Electrodynamics}}},\ Advanced book program\ (\bibinfo
  {publisher} {Westview Press},\ \bibinfo {year} {1998})\BibitemShut {NoStop}%
\bibitem [{\citenamefont {Reiner}\ and\ \citenamefont {Radke}(1990)}]{radke}%
  \BibitemOpen
  \bibfield  {author} {\bibinfo {author} {\bibfnamefont {E.~S.}\ \bibnamefont
  {Reiner}}\ and\ \bibinfo {author} {\bibfnamefont {C.~J.}\ \bibnamefont
  {Radke}},\ }\href@noop {} {\bibfield  {journal} {\bibinfo  {journal} {J.
  Chem. Soc.{,} Faraday Trans.}\ }\textbf {\bibinfo {volume} {86}},\ \bibinfo
  {pages} {3901} (\bibinfo {year} {1990})}\BibitemShut {NoStop}%
\bibitem [{\citenamefont {York}\ and\ \citenamefont {Karplus}(1999)}]{karplus}%
  \BibitemOpen
  \bibfield  {author} {\bibinfo {author} {\bibfnamefont {D.~M.}\ \bibnamefont
  {York}}\ and\ \bibinfo {author} {\bibfnamefont {M.}~\bibnamefont {Karplus}},\
  }\href@noop {} {\bibfield  {journal} {\bibinfo  {journal} {The Journal of
  Physical Chemistry A}\ }\textbf {\bibinfo {volume} {103}},\ \bibinfo {pages}
  {11060} (\bibinfo {year} {1999})}\BibitemShut {NoStop}%
\bibitem [{\citenamefont {Marcus}(1956)}]{marcus}%
  \BibitemOpen
  \bibfield  {author} {\bibinfo {author} {\bibfnamefont {R.~A.}\ \bibnamefont
  {Marcus}},\ }\href@noop {} {\bibfield  {journal} {\bibinfo  {journal} {The
  Journal of Chemical Physics}\ }\textbf {\bibinfo {volume} {24}},\ \bibinfo
  {pages} {966} (\bibinfo {year} {1956})}\BibitemShut {NoStop}%
\bibitem [{\citenamefont {Felderhof}(1977)}]{felderhof}%
  \BibitemOpen
  \bibfield  {author} {\bibinfo {author} {\bibfnamefont {B.~U.}\ \bibnamefont
  {Felderhof}},\ }\href@noop {} {\bibfield  {journal} {\bibinfo  {journal} {The
  Journal of Chemical Physics}\ }\textbf {\bibinfo {volume} {67}},\ \bibinfo
  {pages} {493} (\bibinfo {year} {1977})}\BibitemShut {NoStop}%
\bibitem [{\citenamefont {Jadhao}\ \emph {et~al.}(2012)\citenamefont {Jadhao},
  \citenamefont {Solis},\ and\ \citenamefont {Olvera de~la Cruz}}]{shortpaper}%
  \BibitemOpen
  \bibfield  {author} {\bibinfo {author} {\bibfnamefont {V.}~\bibnamefont
  {Jadhao}}, \bibinfo {author} {\bibfnamefont {F.~J.}\ \bibnamefont {Solis}}, \
  and\ \bibinfo {author} {\bibfnamefont {M.}~\bibnamefont {Olvera de~la
  Cruz}},\ }\href@noop {} {\bibfield  {journal} {\bibinfo  {journal} {Phys.
  Rev. Lett.}\ }\textbf {\bibinfo {volume} {109}},\ \bibinfo {pages} {223905}
  (\bibinfo {year} {2012})}\BibitemShut {NoStop}%
\bibitem [{\citenamefont {Remler}\ and\ \citenamefont {Madden}(1990)}]{madden}%
  \BibitemOpen
  \bibfield  {author} {\bibinfo {author} {\bibfnamefont {D.}~\bibnamefont
  {Remler}}\ and\ \bibinfo {author} {\bibfnamefont {P.}~\bibnamefont
  {Madden}},\ }\href@noop {} {\bibfield  {journal} {\bibinfo  {journal}
  {Molecular Physics}\ }\textbf {\bibinfo {volume} {70}},\ \bibinfo {pages}
  {921} (\bibinfo {year} {1990})}\BibitemShut {NoStop}%
\bibitem [{\citenamefont {Ryckaert}\ \emph {et~al.}(1977)\citenamefont
  {Ryckaert}, \citenamefont {Ciccotti},\ and\ \citenamefont
  {Berendsen}}]{shake}%
  \BibitemOpen
  \bibfield  {author} {\bibinfo {author} {\bibfnamefont {J.-P.}\ \bibnamefont
  {Ryckaert}}, \bibinfo {author} {\bibfnamefont {G.}~\bibnamefont {Ciccotti}},
  \ and\ \bibinfo {author} {\bibfnamefont {H.~J.}\ \bibnamefont {Berendsen}},\
  }\href@noop {} {\bibfield  {journal} {\bibinfo  {journal} {Journal of
  Computational Physics}\ }\textbf {\bibinfo {volume} {23}},\ \bibinfo {pages}
  {327 } (\bibinfo {year} {1977})}\BibitemShut {NoStop}%
\bibitem [{\citenamefont {Arfken}\ and\ \citenamefont {Weber}(2005)}]{arfken}%
  \BibitemOpen
  \bibfield  {author} {\bibinfo {author} {\bibfnamefont {G.}~\bibnamefont
  {Arfken}}\ and\ \bibinfo {author} {\bibfnamefont {H.}~\bibnamefont {Weber}},\
  }\href@noop {} {\emph {\bibinfo {title} {Mathematical Methods For
  Physicists}}}\ (\bibinfo  {publisher} {Elsevier},\ \bibinfo {year}
  {2005})\BibitemShut {NoStop}%
\end{thebibliography}
\end{document}